\documentstyle[aps,epsfig,floats]{revtex}

\def\d{\partial}
\def\s{{\sigma}}
\def\e{{\epsilon}}  
\def\k{{ {\bf k} }}
\def\p{{ {\bf p} }}
\def\q{{ {\bf q} }}

\def\w{{\omega}}
\def\a{{\alpha}}
\def\b{{\beta}}

\def\i{{ {\rm i} }}

\def\I{{ \mbox{\scriptsize I} }}
\def\II{{ \mbox{\scriptsize II} }}
\def\III{{ \mbox{\scriptsize III} }}
\def\IV{{ \mbox{\scriptsize IV} }}
\def\dalt{{ {\raisebox{-0.35ex}[-1.5ex]
 {$\;\stackrel{\leftrightarrow}{\partial}\!\!\!\;$}} }}
\def\eqq{{ {\raisebox{-0.35ex}[-1.5ex]
 {$\;\stackrel{{\mbox{\tiny RTA}}}{=}\;$}} }}

\begin{document}
\draft

\def\runtitle{
General Formula for the Thermoelectric Transport Phenomena
based on the Fermi Liquid Theory
}
\def\runauthor
 {Hiroshi {\sc Kontani}}

\title{
General Formula for the Thermoelectric Transport Phenomena\\
based on the Fermi Liquid Theory: \\
Thermoelectric Power, Nernst Coefficient, and Thermal Conductivity
}

\author{
Hiroshi {\sc Kontani}
}

\address{
Department of Physics, Saitama University,
255 Shimo-Okubo, Urawa-city, 338-8570, Japan.
}

\date{\today}

\maketitle      

\begin{abstract}
On the basis of the linear response transport theory,
the general expressions for the thermoelectric transport coefficients,
such as thermoelectric power ($S$), Nernst coefficient ($\nu$), 
and Thermal conductivity ($\kappa$),
are derived by using the Fermi liquid theory.
The obtained expression is 
exact as for the most singular term in terms of
$1/\gamma_\k^\ast$ 
($\gamma_\k^\ast$ being the quasiparticle damping rate).
We utilize the Ward identities for the heat velocity 
which is derived by the local energy conservation law.
The derived expressions enable us to
calculate various thermoelectric transport coefficients 
in a systematic way, within the framework of the conserving
approximation as Baym and Kadanoff.
Thus, the present expressions are very useful for studying the 
%enables us to study the thermal transport coefficient in 
strongly correlated electrons
such as high-$T_{\rm c}$ superconductors, organic metals,
and heavy Fermion systems,
where the current vertex correction (VC) is expected to 
play important roles.
By using the derived expression, 
we calculate the thermal conductivity $\kappa$ 
in a free-dispersion model up to the second-order 
with respect to the on-site Coulomb potential $U$.
We find that
it is slightly enhanced due to the VC
for the heat current,
although the VC
for electron current makes the conductivity ($\s$) 
of this system diverge,
reflecting the absence of the Umklapp process.

\end{abstract}


%\vskip 5mm
%\hskip 2cm
%keywords: thermoelectric power, Nernst coefficient, thermal conductivity,
%
%\hskip 3cm
%linear response theory, vertex corrections, Fermi liquid theory
%
%\vspace{3mm}

%\begin{multicols}{2}
%\narrowtext

%%%%%%%%%%%%%%%%%%%%%%%%%
% Introduction
%%%%%%%%%%%%%%%%%%%%%%%%%72.10.Bg   
\section{Introduction}
In general,
the transport phenomena in metals 
are very important physical objects
because they offer much information on many-body electronic
properties of the system.
Especially in strongly correlated electron systems
like high-Tc cuprates or heavy Fermion systems,
various transport coefficients show striking
non-Fermi liquid type behaviors.
Historically, theoretical studies on transport phenomena
give considerable progresses 
in various fields of the condensed matter physics,
such as Kondo problem and high-Tc superconductivity.

According to the linear response theory
 \cite{Luttinger,Mahan,Jonson}
or the Kubo formula
 \cite{Kubo},
transport coefficients are given by corresponding
current-current correlation functions.
Thus, to study transport phenomena,
we need to calculate the two-body green function 
with appropriate vertex corrections (VC's).
Unfortunately, in many cases it is a difficult analytical
or numerical work.
Therefore, at the present stage, transport coefficients
are usually studied within the relaxation time approximation
(RTA), by dropping all the VC's.
The effect of the VC can be included
by the standard variational method by Ziman
based on the Boltzmann transport theory
 \cite{Ziman}
However, it is not so powerful for anisotropic correlated systems
because there is no systematic way of choosing the trial function.
Thus, it is desired to establish the microscopic transport theory
based on the linear response formula.

Based on the Kubo formula, 
Eliashberg derived a general expression for 
the dc-conductivity ($\sigma$)
in the Fermi liquid by taking the VC's into account,
by performing the analytic continuation of the current-current
correlation functions
 \cite{Eliashberg}.
Based on the expression, 
Yamada and Yosida proved rigorously that
$\s$ diverges even at finite temperatures
if the Umklapp scattering process is absent.
 \cite{Yamada}.
By generalizing the Eliashberg's theory 
including the outer magnetic field,
the exact formulae for
the Hall coefficient ($R_{\rm H}$)
 \cite{Kohno} 
and the magnetoresistance ($\Delta\rho/\rho$)
 \cite{MR}
in Fermi liquid systems were derived.
By using these formulae, in principle,
we can perform the conserving approximation for these coefficients
with including appropriate VC's 
for currents
 \cite{Baym}.

In general, 
the VC's for currents are expected to be important
especially in strongly correlated systems.
For example, in high-Tc cuprates,
the so-called Kohler's rule 
($|R_{\rm H}|\propto \rho^0$ and $\Delta\rho/\rho\propto \rho^{-2}$)
is strongly violated
 \cite{Kimura}.
%Instead, the relations $\rho\propto T$,
%$|R_{\rm H}|\propto T^{-1}$ and 
%$\Delta\rho/\rho\propto T^{-4}$ are observed.
Moreover, $R_{\rm H}<0$ in electron-doped compounds,
although the shape of the Fermi surface is everywhere hole-like.
These behaviors, which cannot be explained within
the relaxation time approximation (RTA),
had been an open problem in high-Tc cuprates.
Based on the conserving approximation,
we found that these anomalies are well reproduced by the 
VC's for electron currents 
 \cite{Kontani,Kanki,MR-HTSC}.
The effect of the VC's, which are dropped in the RTA,
becomes much important in a Fermi liquid with
strong antiferromagnetic or superconducting fluctuations.

However,
as for the thermoelectric transport coefficients 
such as the thermoelectric power (TEP, $S$), 
the Nernst coefficient ($\nu$),
and the thermal conductivity ($\kappa$),
we do not know
useful expressions for analysis in the 
strongly correlated systems so far.
%have not been derived so far.
Here, the definition of 
$S$,  $\kappa$, and 
$\nu$ under the magnetic filed $B$
parallel to the $z$-axis
are given by
\begin{eqnarray}
S &=& -E_x/\d_x T,
 \nonumber \\
\kappa &=& -Q_x/\d_x T,
 \\
\nu &=& -E_y/B\d_x T,
 \nonumber
\end{eqnarray}
where ${\vec Q}$ is the heat current.
Unfortunately,
the conserving approximation 
for these coefficients is not practicable
because we do not know how to calculate the VC's 
for them.
Thus, at the present stage, 
only the RTA is accessible,
which will be insufficient
%a reliable analysis on the thermoelectric transport coefficients
for a reliable analysis of the strongly correlated systems
because VC's should be included.

There is a long history of the microscopic study on 
thermoelectric transport phenomena.
To this problem, we cannot apply the Kubo formula to
the electronic conductivity naively
because there is no Hamiltonian
which describes a temperature gradient $\d_x T$.
In 1964 Luttinger gave a microscopic proof
%mechanical proof
that thermoelectric transport coefficients
are given by the corresponding 
current-current correlation function
 \cite{Luttinger}.
Later, Mahan et al.
much simplified the Luttinger's expression
%and derived expressions for transport coefficients  
%in terms of the Green function,
in the case of electron-phonon and electron-impurity
interactions
 \cite{Jonson}.
However, the analysis of the VC for the heat current
for electron-electron interacting systems
is still an open problem, which is necessary
to go beyond the RTA.
This analysis will be more complicated
and profound than that for the electric conductivity
performed by Eliashberg
 \cite{Eliashberg}.

In the present paper,
%we present the systematic way of the calculation of the vertex 
%correction for various thermoelectric transport coefficients.
%we derive the general expressions for $S$, $\nu$ and $\kappa$
%by using the Fermi liquid theory.
we derive the thermoelectric transport coefficients
by performing the analytic continuations of the 
current-current correlation functions,
on the basis of the linear response theory 
%for thermoelectric transport phenomena 
developed by Luttinger or Mahan.
Our expressions are valid for general two-body interactions.
The VC for the heat current 
is given without ambiguity by the Ward identity
with respect to the local energy conservation law.
The derived expressions are ``exact'' as for the
most divergent term with respect to $\gamma_\k^{-1}$,
where $\gamma_\k$ is the quasiparticle damping rate.
%Because the obtained expression for $S$, $\nu$ and $\kappa$
%contains all the VC's which ensure the conservation laws,
The present work enables us to 
perform the ``conserving approximation'' 
for $S$, $\nu$ and $\kappa$,
which is highly demanded to avoid unphysical results. 
Actually, the VC's would totally modify
the behavior of these quantities
in strongly correlated electron systems,
as it does for the Hall effect and the magnetoresistance.
In this respect, the RTA is unsatisfactory
%unreliable
because all the current VC's are neglected there.

We note that Langer studied the Ward identity for the heat 
current, and discussed the thermal conductivity
 \cite{Langer}.
However, the derived Ward identity was not correct
as explained in \S III, although it did not influence 
the thermal conductivity at lower temperatures fortunately.
We derive the correct Ward identity
in \S III, and give expressions for $\kappa$, $S$ and $\nu$.

In thermodynamics, the TEP of metals
becomes zero at absolute zero temperature,
which is the consequence of 
``the third law of the thermodynamics''.
(As is well known, the third law also tells that
the heat capacity vanishes at $T=0$.)
Similarly,
both $\nu$ and $\kappa$ also become zero at $T=0$
if the quasiparticle relaxation time $\tau \ (=1/2\gamma)$ 
is finite at $T=0$ due to impurity scatterings.
Unfortunately,
these indisputable facts are nontrivial
in a naive perturbation study
once the electron-electron correlations are set in.
In the present work,
we derive the general expression for $S$
which automatically satisfy $S(T=0)=0$
owing to the Ward identity for the heat velocity.

In high-Tc cuprates,
the Nernst coefficient $\nu$ increases drastically
below the pseudo-gap temperature,
which is never possible to explain
within the RTA
 \cite{Ong}.
According to recent theoretical works,
superconducting (SC) fluctuation is
one of the promising origins of the pseudo-gap phenomena 
 \cite{Levin,Koikegami,Yanase,Kobayashi}.
Based on the opinion,
we studied $\nu$ for high-Tc cuprates 
using the general expression derived in the present paper
 \cite{Nernst-HTSC}.
Then, we could reproduce the rapid increase of $\nu$ 
only when the VC's due to the strong antiferromagnetic and 
superconducting fluctuations are taken into account.
This work strongly suggests that the origin of the pseudo-gap 
phenomena in high-Tc cuprates is the strong 
$d$-wave superconducting fluctuations.

In the case of heavy Fermion systems,
the TEP takes an enhanced value around the coherent temperature,
and its sign changes in some compounds at lower temperatures.
Such an interesting non-Fermi liquid like behavior 
is mainly attributed to a huge energy dependence
of the relaxation time, $\tau(\e) \ (=1/2\gamma(\e))$, 
due to the Kondo resonance.
This phenomenon was studied by using the
dynamical mean field theory
\cite{Czy,Kotliar}.
Also, the TEP in the Kondo insulator was studied  
in ref. \cite{Saso} in detail.

The contents of this paper are as follows:
In \S II, we develop the linear response theory
for thermoelectric transport coefficients.
By performing the analytic continuation,
we derive the general formula of $S$ and $\kappa$ 
in the presence of the on-site Coulomb potential $U$.
In \S III, we derive the Ward identity 
for the heat velocity which is valid 
for general two-body interactions, 
by using the local energy conservation law.
The Ward identity assures that
the expressions for $S$ and $\kappa$ 
derived in the previous section are valid 
even if the range of the interaction is finite,
as for the most divergent terms with respect to $\gamma^{-1}$.
%Another derivation based on the diagrammatic 
%technique is explained in Appendix C.
In \S IV, the general formula for $\nu$ is derived.
It is rather a complicated task because  
the gauge invariance should be maintained.
In \S V, we calculate $\kappa$ in a spherical correlated
electron system in the absence of Umklapp process,
and obtain its exact expression
by including the VC's within the second order perturbation.
The physical meaning of the VC is discussed.
Finally, the summary of the present work is shortly 
expressed in \S VI.

%%%% punch line %%%
%The aim of this paper is to derive the
%{\it general expression for the thermal transport coefficients}
%from the linear response theory,
%initiated by Luttinger.
%The obtained expression for $S$, $\nu$ and $\kappa$
%contains all the vertex corrections which ensure the conserving laws.
%In the system with strong correlations,
%the vertex corrections will qualitatively change
%the behavior of these quantities,
%as it does for the Hall effect and the magnetoresistance.
%In this respect, the RTA is unreliable
%because all the current vertex corrections are neglected there.
%Based on the derived expression,
%we will study the MR in high-$T_{\r${\vec X}^1\equiv{\vec E}/T$ m c}$ cuprates 
%and discuss the violation of Kohler's rule in later publications
% \cite{Future}.

%%%%%%%%%%%%%%%%%%%%%%%%%%
% 2 Linear response theory
%%%%%%%%%%%%%%%%%%%%%%%%%%
\section{Linear Response Theory for $S$ and $\kappa$}
First,
we shortly summarize the 
linear response theory for thermoelectric transport coefficients,
initiated by Luttinger.
Here we consider the situation that both 
the electron current ${\vec J}^1\equiv{\vec J}$ 
and the heat current ${\vec J}^2\equiv{\vec Q}$
are caused by the external forces
${\vec X}^1\equiv{\vec E}/T$ and 
${\vec X}^2\equiv{\vec \nabla}\left(\frac1T\right)$,
where ${\vec E}$ is the electric field.
In the linear response,
\begin{eqnarray}
{\vec J}^l = \sum_{m=1,2}{\hat L}^{lm}({\vec B}){\vec X}^m,
\end{eqnarray}
where $l,m=1,2$.
Because the relation 
$\frac{dS_e}{dt}=\sum_l{\vec J}^l\cdot {\vec X}^l$
($S_e$ being the entropy) is satisfied 
in the present definition, the tensor 
${\hat L}^{lm}({\vec B})$ satisfy the Onsager relation;
$L_{\mu\nu}^{lm}({\bf B}) = L_{\nu\mu}^{ml}(-{\bf B})$,
where $\mu,\nu=x,y,z$
 \cite{deGoodt}.

According to the quantum mechanics,
the electron current operator ${\vec j}$ and 
the heat current one ${\vec j}^Q$ are given by
 \cite{Mahan}
\begin{eqnarray}
{\vec j}({\bf r}_i) &=& i[H, e\rho({\bf r}_i){\bf r}_i],
 \label{eqn:def-j} \\
{\vec j}^Q({\bf r}_i) 
 &=& {\vec j}^E({\bf r}_i) -\frac{\mu}{e} {\vec j}({\bf r}_i),
 \label{eqn:def-jq} \\
{\vec j}^E({\bf r}_i) &=& i[H,h({\bf r}_i){\bf r}_i],
\end{eqnarray}
where $e (<0)$ is the charge of an electron, 
$H$ is the Hamiltonian without external fields $X^l$, 
$\rho({\bf r}_i)=\sum_\s c_\s^\dagger({\bf r}_i) c_\s({\bf r}_i)$
is the density operator
($\s$ being the spin suffix), and
$h({\bf r}_i)$ is the local Hamiltonian
by which $H$ is given by
$H= \sum_{i}h({\bf r}_i)$.
By using these current operators,
${\vec J}$ and ${\vec Q}$ are given by
${\vec J}({\bf r}_i)= \langle {\vec j}({\bf r}_i) \rangle$ and
${\vec Q}({\bf r}_i)= \langle {\vec j}^Q({\bf r}_i) \rangle$,
respectively.

To derive the expressions for 
various conductivities microscopically, 
we introduce the virtual external potential term $F$ 
which causes the currents $X^l$ $(l=1,2$).
Then, the ``total Hamiltonian'' is expressed as
$H_{\rm T}= H + F \cdot e^{(-i\w+\delta)t}$, where $\delta>0$ is 
a infinitesimally small constant.
According to the linear response theory
 \cite{Luttinger,Kubo},
the current $J^l$ at $t$ is given by
\begin{eqnarray}
{\vec J}^l(t) 
 &=& \left\langle {\vec j}^l({\q}=0,t) \right\rangle
 \nonumber \\
 &=& -i\int_{-\infty}^t dt' \left\langle 
 [{\vec j}^l({\q}=0,t),F(t')] \right\rangle e^{(-i\w+\delta)t'}.
\end{eqnarray}
Because of the relation
$\displaystyle \frac{\d F}{\d t} = T\frac{\d S_{\rm e}}{\d t}
 = T\sum_l {\vec j}^l\cdot {\vec X}^l$
the expression for ${\hat L}^{lm}$ is given by
 \cite{Mahan}
\begin{eqnarray}
{\hat L}^{lm} &=& 
 \left.{\hat L}^{lm}(\w+\i\delta)\right|_{\w=0} ,
 \label{eqn:def-L} \\
{L}_{\mu\nu}^{lm}(\i\w_l) &=& \frac{-T}{\w_l} \int_0^\beta 
 e^{i\w_l\tau} \left\langle T_\tau j_\mu^l({\q}=0,\tau)
 j_\nu^m({\q}=0,\tau=0) \right\rangle ,
 \label{eqn:def-K}
\end{eqnarray}
where $\beta=1/T$ and $\mu,\nu=x,y$.
$T_\tau$ is the $\tau$-ordering operator, and
$\w_l= 2\pi T l$ ($l$ being the integer) is the bosonic 
Matsubara frequency.
By writing the diagonal component of ${\hat L}^{lm}$
as ${L}^{lm}$,
$\s$, $S$ and $\kappa$ are given by
 \cite{Mahan}
\begin{eqnarray} 
\s  &=& \frac{e^2}{T}L^{11},
 \\
S   &=& \frac1{eT}\frac{L^{21}}{L^{11}} 
 = \frac{e}{T^2} \frac{L^{21}}{\s}, 
 \label{eqn:S-L12}
 \\
\kappa &=& \frac1{T^2}\left(  
 L^{22}-\frac{L^{12}L^{21}}{L^{11}} \right) ,
 \label{eqn:K-L22} 
\end{eqnarray} 
where $e$ $(e<0)$ is the charge of an electron.

Hereafter, we analyze the function ${\hat L}^{lm}(\i\w_l)$
given by eq.(\ref{eqn:def-K}) at first, 
and perform the analytic continuation to derive
${\hat L}^{lm}$ by eq.(\ref{eqn:def-L}).
We study a tight-binding model with two-body interactions,
which is expressed in the absence of the magnetic field
as:
\begin{eqnarray}
H &=& H_0+H_{\rm int},
 \label{eqn:Hubbard} \\
H_0 &=& \sum_{\k,\s}\e_\k^0 c_{\k\s}^\dagger c_{\k\s} ,
 \label{eqn:H0}
 \\
H_{\rm int} &=& \frac12 \sum_{\k\k'\q\s\s'} U_{\s\s'}(\q) 
 c_{\k+\q,\s}^\dagger c_{\k'-\q,-\s}^\dagger c_{\k',-\s} c_{\k,\s}.
 \label{eqn:Hint} 
\end{eqnarray}
In eq.(\ref{eqn:H0}), $\e_\k^0 = \sum_{i} t_{i,0}
 e^{i{\k}\cdot({\bf r}_i-{\bf r}_0)}$,
where $t_{ij}$ is the hopping parameter between
${\bf r}_i$ and ${\bf r}_j$.
$U_{\s\s'}(\q)$ represents the electron-electron
correlation between $\s$ and $\s'$ spins.
For example, $U_{\s\s'}(\q) \equiv U\delta_{\s,-\s'}$
for the on-site Coulomb interaction.

The one-particle Green function is given by
$G_\k(\e)= 1/(\e+\mu-\e_\k^0 - \Sigma_\k(\e))$,
where $\Sigma_\k(\e)$ is the self-energy
and $\mu$ is the chemical potential.
In a Fermi liquid,
$\gamma_\k \ll T$ is satisfied
at sufficiently low temperatures
because of the relation $\gamma_\k \propto T^2$
 \cite{AGD}.
In such a temperature region,
the following quasiparticle representation 
of the Green function is possible:
\begin{eqnarray}
G_\k(\w) = \frac{z_\k}{\w-\e_\k^\ast + i\gamma_\k^\ast},
 \label{eqn:Green-spect}
\end{eqnarray}
%
%In eq.(\ref{eqn:Green-spect}), 
where
$z_\k=(1-\frac{\d}{\d\w}\Sigma_\k(\e))^{-1}$ 
is the renormalization factor,
$\e_\k^\ast= z_\k(\e_\k^0+\Sigma_\k(0)-\mu)$,
$\gamma_\k^\ast= z_\k \gamma_\k$ and
$\gamma_\k= {\rm Im}\Sigma_\k(-i0)$,
respectively.

According to eq.(\ref{eqn:def-j}),
the electron current operator for eq.(\ref{eqn:Hubbard}) 
is given by
\begin{eqnarray}
{\vec j}(\p) = e\sum_{\k,\s}{\vec v}_{\k}^0 
 c_{\k-\p/2,\s}^\dagger c_{\k+\p/2,\s}, 
 \label{eqn:vero-q}
\end{eqnarray}
where ${\vec v}_{\k}^0 = {\vec \nabla}_k \e_\k^0$.
Apparently, ${\vec j}(\p)$ is a one-body operator.

In the same way, we consider 
the heat current operator defined by eq.(\ref{eqn:Hubbard}):
In the case of the on-site Coulomb interaction,
for simplicity, it is obtained after a long but 
straightforward calculation as
\begin{eqnarray}
{\vec j}^Q(\p=0) &=& \sum_{\k,\s}(\e_\k^0-\mu){\vec v}_\k^0
 c_{\k\s}^\dagger c_{\k\s}
 \nonumber \\
&+& \frac U2\sum_{\k\k'\q\s}
 \frac12({\vec v}_{\k+\q/2}^0+{\vec v}_{\k-\q/2}^0)
 c_{\k-\q/2,\s}^\dagger c_{\k+\q/2,\s} 
 c_{\k'+\q/2,-\s}^\dagger c_{\k'-\q/2,-\s} ,
 \label{eqn:jq}
\end{eqnarray}
which contains a two-body term in the case of $U\ne0$.
It becomes more complicated
for general long-range potential.
This fact seems to make the analysis of the
thermoelectric coefficient very difficult.

%The derivation of eq.(\ref{eqn:jq}) derivation is given 
%in Appendix.

Fortunately,
as shown in Appendix A,
eq.(\ref{eqn:jq}) can be transformed into 
the following simple one-body operator form
by using the kinetic equation:
\begin{eqnarray} 
{\vec j}^Q(\p=0,\w_l)
&=& \sum_{\k,\s}\int_0^{\b} d\tau e^{i\w_l\tau}
 \lim_{\tau'\rightarrow\tau} 
 \frac12 \left( \frac{\d}{\d\tau}-\frac{\d}{\d\tau'} \right)
 c_{\k,\s}^\dagger(\tau) 
 c_{\k,\s}(\tau') 
 \nonumber \\
&=& T\sum_{\k\e_n\s} i(\e_n+\w_l/2){\vec v}_\k^0 \cdot 
 c_{\k,\s}^\dagger(\e_n) c_{\k,\s}(\e_n+\w_l) ,
 \label{eqn:jqM}
\end{eqnarray}
where $\w_l$ and $\e_n$ are boson and fermion Matsubara
frequencies, respectively.
The case of the nonlocal electron-electron interaction
is discussed in the next section
by constructing the Ward identity for the heat velocity.
%More noteworthy are,
%according to the Ward identity for the heat current 
%which will be discussed in the next section 
%(eq.(\ref{eqn:q-bare})),
%the bare heat current is expressed as eq.(\ref{eqn:jqM})
%for general correlated electron systems.

\begin{figure}
\begin{center}
\epsfig{file=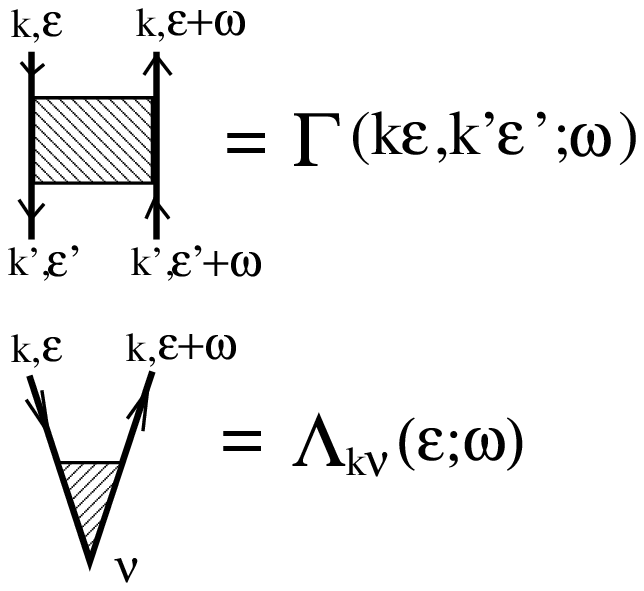,width=4cm}
\end{center}
\caption{
The four-point vertex correction $\Gamma$
and the three-point vertex function $\Lambda_\nu$,
respectively.
}
  \label{fig:34VCs}
\end{figure}

By using eq.(\ref{eqn:jqM}),
we obtain the expression for $L^{12}$ 
without the magnetic field as follows:
\begin{eqnarray} 
L_{\mu\nu}^{12}(i\w_l) 
 &=& \frac{T^2 e}{\w_l}\sum_{\k,\e_n} i (\e_n+\frac12\w_l)v_{\k\mu}^0
  g_\k(\e_n,\w_l) \Lambda_{\k\nu}(\e_n,\w_l) ,
 \label{eqn:L12M} \\
\Lambda_{\k\nu}(\e_n,\w_l)
 &=& v_{\k\nu}^0
 + T\sum_{\k',\e_{n'}}
  \Gamma(\k\e_n,\k'\e_{n'};\w_l)
  g_{\k'}(\e_{n'},\w_l) v_{\k'\nu}^0,
 \label{eqn:L12-Matsubara}
\end{eqnarray} 
where 
$g_\k(\e_n,\w_l) \equiv G_\k(i\e_n+i\w_l) G_\k(i\e_n)$.
$\Lambda_{\k\nu}(\e_n,\w_l)$ and
$\Gamma(\k\e_n,\k'\e_{n'};\w_l)$
are the three- and four-point vertices respectively,
which are expressed in Fig.
 \ref{fig:34VCs}. 
They are reducible with respect to
the particle-hole channel.
Note that we put the outer momentum $\p=0$ 
in eq.(\ref{eqn:L12-Matsubara})
because we are interested in the dc-conductivity. 

%%%%%%%%%%%%%%%%%%%%%%%%%%%%%%%%%%%%%%%%%%%%%%%%%%%%%%
%%%%%%%%%%%%%%%%%%%%%%%%%%%%%%%%%%%%%%%%%%%%%%%%%%%%%%

The expression (\ref{eqn:L12M}) for $L^{12}$
derived for the on-site Coulomb interaction
is equal to that for a system with the impurity scattering
and the electron-phonon interaction
derived by Jonson and Mahan.
 \cite{Jonson}.
In the next section, we will show that
the expression is also valid for general types of 
two-body interactions as for the most divergent term
with respect to $\gamma^{-1}$
on the basis of the Ward identity.
%The correction for the heat current
%due to the ``momentum derivative of $U(\q)$''
%turns out to give negligible contributions
%for $S$, $\nu$ and $\kappa$. (see Appendix C.)

%In the same way, $L_{\mu\nu}^{22}$ is given as
%%
%\begin{eqnarray} 
%L_{\mu\nu}^{22}(i\w_l) 
% &=& + \frac{T}{i\w_l}\sum_{k,n} (\e_n+\frac12\w_l)^2 v_{k\mu}^0
%  g_k(\e_n,\w_l) v_{k\nu}^0
%   \nonumber \\
% & &+ \frac{T^2}{i\w_l}\sum_{kk',nn'} (\e_n+\frac12\w_l) v_{k\mu}^0
%  g_k(\e_n,\w_l)\Gamma_{kk'q}(\e_n,\e_{n'};\w_l)
%  g_{k'}(\e_{n'},\w_l) (\e_{n'}+\frac12\w_l) v_{k'\nu}^0.
%\end{eqnarray} 
%%

The dc-TEP is obtained by the analytic continuation 
of eq.(\ref{eqn:L12M}) with respect to $i\w_l$,
by taking all the VC's into account.
(The analysis on the VC's in ref.\cite{Jonson}
is insufficient.)
In the present work, we perform the analytic continuation
rigorously by referring to the Eliashberg's procedure
in ref.\cite{Eliashberg}.
Next, by using the Ward identity (\ref{eqn:Ward}),
we derive the simple expression for the
TEP within the 
most divergent term with respect to $\gamma^{-1}$.

After the analytic continuation of eq.(\ref{eqn:L12M}),
$L_{\mu\nu}^{12}(+i\delta)$
of order $O(\gamma^{-1})$ is given by
\begin{eqnarray} 
L_{\mu\nu}^{12}(+i\delta)
 &=& eT \sum_\k \int \frac{d\e}{\pi} \left(-\frac{\d f}{\d \e}\right)
 q_{\k\mu}(\e)|G_\k(\e)|^2 J_{\k\nu}(\e),
  \label{eqn:L12} 
%\left. \frac1{\i\w} L_{\mu\nu}^{22}(\w+i\delta)\right|_{\w=0}
% &=& \sum_\k \int \frac{d\e}{\pi} \left(-\frac{\d f}{\d \e}\right)
% q_{k\mu}(\e)|G_k(\e)|^2 Q_{k\nu}(\e),
%  \label{eqn:L22}
\end{eqnarray}
where the total electron current $J_{\k\nu}$ with VC's
and the quasiparticle heat velocity $q_{\k\nu}$ are 
respectively given by
\begin{eqnarray} 
{\vec J}_{\k}(\e)
 &=&  {\vec v}_{\k}(\e) 
 + \sum_{\k'} \int \frac{d\e'}{4\pi i} \left(-\frac{\d f}{\d \e}\right)
 {\cal T}^{22}(\k\e,\k'\e') g_{\k'}^{(2)}(\e') {\vec v}_{\k'}(\e'),
 \label{eqn:J}
  \\
{\vec v}_\k(\e) 
 &=& {\vec v}_\k^0 + \sum_{\k',i=1,3} \int \frac{d\e'}{4\pi i}
 {\cal T}^{2i}(\k\e,\k'\e') g_{\k'}^{(i)}(\e') {\vec v}_{\k'}^0,
%{\vec \nabla}_k{\rm Re}\Sigma_k(\e)
  \label{eqn:v}
  \\
{\vec q}_{\k}(\e)
 &=& \e {\vec v}_\k^0
 + \sum_{\k',i=1,3} \int \frac{d\e'}{4\pi i}
 {\cal T}^{2i}(\k\e,\k'\e') g_{\k'}^{(i)}(\e') \e' {\vec v}_{\k'}^0,
  \label{eqn:q}
\end{eqnarray}
where $g_\k^{(1)}(\e) = \{G_\k^{\rm A}(\e)\}^2$, 
$g_\k^{(2)}(\e) = |G_\k(\e)|^2$ and 
$g_\k^{(3)}(\e) = \{G_\k^{\rm R}(\e)\}^2$, respectively.
The definition of the four-point vertex
${\cal T}^{m,l}(\p\e,\p'\e')$ is given
in ref. \cite{Eliashberg},
which are listed in Appendix B.
In general,
${\cal T}^{2i}$ is 
well approximated at lower temperatures as
$({\cal T}^{1i}+{\cal T}^{3i})/2$
 \cite{Eliashberg}.
Thus, taking the Ward identity
for electron current is taken into account
 \cite{AGD,Nozieres},
${\vec v}_{\k}(\e)$ is simply given by
\begin{eqnarray}
{\vec v}_{\k}(\e)
 = {\vec \nabla}_k(\e_\k^0+{\rm Re}\Sigma_\k(\e)) .
  \label{eqn:v2}
\end{eqnarray}

Next, we consider ${\vec q}_{\k}(\e)$
defined in eq.(\ref{eqn:q}):
By seeing its functional form,
the relation ${\vec q}_{\k}(\e=0) =0$
is nontrivial.
However, 
if ${\vec q}_{\k}(\e=0)$ were nonzero,
then $L^{12}$ in eq.(\ref{eqn:L12M})
would be proportional to $T \gamma^{-1}$.
In this case,
$S = eL^{12}/\s T^2$ diverges at $T=0$,
which contradicts 
``the third law of the thermodynamics''!
In this sense, eq.(\ref{eqn:q}) is too primitive
for a reliable (numerical) analysis
at lower temperatures.

Fortunately, 
by noticing that
${\cal T}^{2i} = ({\cal T}^{1i}+{\cal T}^{3i})/2$
at lower temperatures,
the quasiparticle heat velocity ${\vec q}_{\k}(\e)$ 
given in eq.(\ref{eqn:q}) can be rewritten in a simple form as
\begin{eqnarray}
{\vec q}_\k(\e) &=& \e {\vec v}_\k(\e) ,
% \nonumber \\ 
%&=&  \e({\vec v}_k^0 
% + {\vec \nabla}_k{\rm Re}\Sigma_\k(\e) )
 \label{eqn:Ward}
\end{eqnarray}
where ${\vec v}_\k(\e)$ is given in eq.(\ref{eqn:v2}).
Equation (\ref{eqn:Ward}) is the Ward identity 
which will be derived from the local energy conservation
law in the next section.
This Ward identity 
leads to $L^{12}\propto T^3\gamma^{-1}$
because of ${\vec q}_{\k}(\e=0) =0$,
so the difficulty in analyzing 
the TEP towards $T\rightarrow 0$ is removed.

In the same way, we derive the exact formula for the
thermal conductivity within the most divergent term 
with respect to $\gamma^{-1}$.
By the similar way to the derivation of eq.(\ref{eqn:L12M}),
we obtain that
\begin{eqnarray} 
L_{\mu\nu}^{22}(i\w_l) 
 &=& \frac{T^2}{\w_l}\sum_{\k,\e_n} i (\e_n+\frac12\w_l)v_{\k\mu}^0
  g_\k(\e_n,\w_l) \Lambda_{\k\nu}^Q(\e_n,\w_l) ,
 \label{eqn:L22M} \\
\Lambda_{\k\nu}^Q(\e_n,\w_l)
 &=& i(\e_{n}+\frac12\w_l) v_{\k\nu}^0
 + T\sum_{\k',\e_{n'}}
  \Gamma(\k\e_n,\k'\e_{n'};\w_l)
  g_{\k'}(\e_{n'},\w_l) \cdot i(\e_{n'}+\frac12\w_l) v_{\k'\nu}^0 .
 \label{eqn:L22-Matsubara}
\end{eqnarray} 
%
%where 
%${q}_{\k\nu}^0(\e_n,\w_l)\equiv i(\e_{n}+\frac12\w_l) v_{\k\nu}^0$.
After the analytic continuation,
we find that
\begin{eqnarray}  
L_{\mu\nu}^{22}(+i\delta)
 &=& T\sum_\k \int \ \frac{d\e}{\pi} \left(-\frac{\d f}{\d \e}\right)
 q_{\k\mu}(\e)|G_\k(\e )|^2 Q_{\k\nu}(\e),
  \label{eqn:L22} \\
{\vec Q}_{\k}(\e) 
 &=&   {\vec q}_{\k }(\e)  
 + \sum_{\k'} \int \frac{d\e'}{4\pi i} \left(-\frac{\d f}{\d \e}\right)
 {\cal  T}^{22}(\k\e,\k'\e') g_{\k'}^{(2)}(\e') {\vec q}_{\k'}(\e') ,
 \label{eqn:Q}
\end{eqnarray}
where ${\vec Q}_{k}(\e)$ is the total heat current with VC's,
which is given by the analytic continuation of 
${\Lambda}_{\k x}^Q(i\e_n;i\w_l)$
from the region $\e_n+\w_l>0$ and $\e_n<0$,
and by taking the limit $\w\rightarrow 0$.
We stress that ${\vec Q}_\k(\e=0)=0$
at zero temperature,
because the $\e'$-integration range in eq.(\ref{eqn:Q})
is restricted to within $|\e'|\sim T$ 
due to the thermal factors in ${\cal T}^{22}$
given by the analytic continuation;
see Appendix B.
This fact leads to the result that $L_{\mu\nu}^{22}(+i\delta)$
given by eq.(\ref{eqn:L22}) is proportional to $T^3\gamma^{-1}$
as $T\rightarrow0$.
As a result, at lower temperatures,
the relation $\kappa \propto T\gamma^{-1}$ is assured
in the present analysis.

In summary,
(i) the TEP is given by
$\displaystyle S= \frac{e}{T^2\s}{L^{12}}$,
where 
$L^{12}$ is given in eq.(\ref{eqn:L12})
and ${\vec J}$, ${\vec v}$ and ${\vec q}$ are
given by eq.(\ref{eqn:J}), eq.(\ref{eqn:v2}) and
eq.(\ref{eqn:Ward}), respectively.
(ii) the thermal conductivity is given by
$\displaystyle \kappa = \frac{L^{22}}{T^2} - T S^2\s$,
where $L^{22}$ is given in eq.(\ref{eqn:L22}).
At lower temperatures,
the first term $L^{22}/T^2$ is dominant 
because it is proportional to $T\gamma^{-1}$
whereas the second term is proportional to $T^3\gamma^{-1}$.

Finally, we note that the conductivity $\s$ is given by
 \cite{Eliashberg}
\begin{eqnarray}  
\s_{\mu\nu} = e^2 \sum_\k \int \ \frac{d\e}{\pi} 
 \left(-\frac{\d f}{\d \e}\right)
 v_{\k\mu}(\e)|G_\k(\e )|^2 J_{\k\nu}(\e),
  \label{eqn:sigma} 
\end{eqnarray} 
where ${\vec J}_\k(\e)$ is given by eq.(\ref{eqn:J}).

%%%%%%%%%%%%%%%%%%%%%%%%%%%%%%%%%%%%%%%%
% Ward identity
%%%%%%%%%%%%%%%%%%%%%%%%%%%%%%%%%%%%%%%%
\section{Generalized Ward Identity}
As we discussed in \S I,
the TEP becomes zero at zero temperature,
which are ensured by ``the third law of the thermodynamics''.
This fact means that $L^{12}$ given by eq.(\ref{eqn:L12})
% and eq.(\ref{eqn:L22})
should be proportional to $T^3\gamma^{-1}$ as $T\rightarrow0$.
This relation is ensured if the Ward identity 
(\ref{eqn:Ward}) is satisfied exactly.
In this section,
we derive the generalized Ward identity 
for general types of the two-body interactions,
by noticing the local energy conservation law;
$\frac{\d}{\d t}h + {\vec \nabla}\cdot{\vec j}^Q=0$,
where $h(z)$ is the local part of the Hamiltonian.
The obtained heat velocity has a correction term 
$\Delta {\vec q}$ (see eq.(\ref{eqn:qi2-final})),
which turn out to be negligible for transport coefficients
at lower temperatures.
The present derivation is analogous to the proof
for the generalized Ward identity for the electron current
which is described in ref.
 \cite{Schrieffer}.
In this section,
we drop the spin suffixes
for simplicity of the description.
In the same reason, we put $\mu=0$
because it will not cause a confusion.

Here, we introduce the four-dimensional heat velocity;
 $(h(z),{\vec j}^Q(z))\equiv (j_0^Q(z),j_1^Q(z),j_2^Q(z),j_3^Q(z))$.
Then, we consider the following function $X_\mu^Q$ ($\mu=0\sim3$):
\begin{eqnarray}
X_\mu^Q(x,y,z)
 &\equiv& \left\langle T_\tau j_\mu^Q(z) c(x)c^\dagger(y) 
  \right\rangle
  \nonumber \\
 &=& \int\!\!\int G(x,x') \Lambda_\mu^Q(x',y',z)G(y',y)d^4x'd^4y' ,
\end{eqnarray}
where $\Lambda_\mu^Q(x',y',z)$ is the
three-point vertex function with respect to the heat velocity.
For the simplicity of the description,
we assume hereafter that $x$, $y$ and $z$ are continuous variables,
not discrete ones.
Because of the translationally invariance of the system, 
we can write as
\begin{eqnarray}
X_\mu^Q(x,y,z)
 &\equiv& \int\!\!\int X_\mu^Q(k+p,k) 
 e^{i(k({x}-{y})+p({x}-{z}))}
% e^{-i(k_0(x_0-y_0)+p_0(x_0-z_0))}
 d^4kd^4p .
 \label{eqn:Fourier}
\end{eqnarray}
Here, we use the following four-dimensional notations
for $x$, $p$ and $k$;
$(t,x_1,x_2,x_3)\equiv x_\mu$,
$(-\w,p_1,p_2,p_3)\equiv p_\mu$, and
$(-\e,k_1,k_2,k_3)\equiv k_\mu$.

For the moment,
we assume that $h(z)$ is a local operator
i.e., the two-body interaction $U({\bf x}-{\bf y})$
is a $\delta$-function type.
(This restriction on $h(z)$ is released
later in the present section.)
Because the relation $i[h(z),c(x)]\delta(z_0-x_0)
= \frac{\d}{\d x_0}c(x)\delta^4(z-x)$
is satisfied, then
\begin{eqnarray}
\sum_{\nu=0}^3 \frac{\d X_\nu}{\d z_\nu} 
%  + \frac{\d X_0}{\d z_0}
&=& \left( T_\tau 
 \left( \sum_{\nu=0}^3 \frac{\d j_\mu^Q(z)}{\d z_\nu}
% + \frac{\d j_0^Q(z)}{\d z_0}
  \right) c(x)c^\dagger(y) \right\rangle
  \nonumber \\
& &+ \left\langle T_\tau [h(z),c(x)] c^\dagger(y) 
     \right\rangle \delta(z_0-x_0)
  \nonumber \\
& &+ \left\langle T_\tau c(x) [h(z),c^\dagger(y)] 
     \right\rangle \delta(z_0-y_0)
  \nonumber \\
&=& \frac{\d}{\d x_0}G(x-y) \cdot \delta^4(z-x)
  + \frac{\d}{\d y_0}G(x-y) \cdot \delta^4(y-z)  ,
 \label{eqn:X4}
\end{eqnarray}
where $\delta^4(x)\equiv\delta({\bf x})\delta(x_0)$.
In the transformation,
the local energy conservation law is taken into account.
Performing the Fourier transformation of eq.(\ref{eqn:X4}),
we get
\begin{eqnarray}
%-p_0 \Lambda_0^Q(k+p,k)
\sum_{\nu=0}^3 p_\nu \Lambda_\nu^Q(k+p,k)
 = (\e_{\bf k}^0 + \Sigma(k))(k_0+p_0)
   -(\e_{\bf k+p}^0 + \Sigma(k+p)) k_0 .
 \label{eqn:X4-2}
\end{eqnarray}
%
%The solution of the above equation is given by
%%
%\begin{eqnarray}
%\Lambda_i^Q(k+p,k)
% &=& \frac{p_i}{|\p|^2} 
% (\e_{\bf k+p}^0 + \Sigma(k+p) - \e_{\bf k}^0 - \Sigma(k))
% \cdot (-k_0-p_0/2) ,
% \label{eqn:Lqi} \\
%\Lambda_0^Q(k+p,k) 
% &=& \frac1{2} 
% (\e_{\bf k+p}^0 + \Sigma(k+p) + \e_{\bf k}^0 + \Sigma(k)) ,
% \label{eqn:Lq0} 
%\end{eqnarray}
%%
%where $i=1,2,3$.
%Equations (\ref{eqn:Lqi}) and (\ref{eqn:Lq0})
%are the generalized Ward identity.
%By taking the limit $\p\rightarrow 0$ of 
%eq.(\ref{eqn:X4}),
%we obtain the bare heat current (without VC's) as
%%
%\begin{eqnarray}
%\Lambda_i^{Q0}(\k\e+\w,\k\e)= (\e+\w/2) \cdot {\vec v}_\k^0 .
% \label{eqn:q-bare}
%\end{eqnarray}
%%
%Thus, we find that the heat current operator
%is given by eq.(\ref{eqn:jqM}).

%Even so, a careful numerical calculation 
%is required to derive a reliable 
%result at lower temperatures.
%In these respects,
%the Ward identity (\ref{eqn:Ward}) is very useful
%for studying $L^{12}$ and $L^{22}$.

\begin{figure}
\begin{center}
\epsfig{file=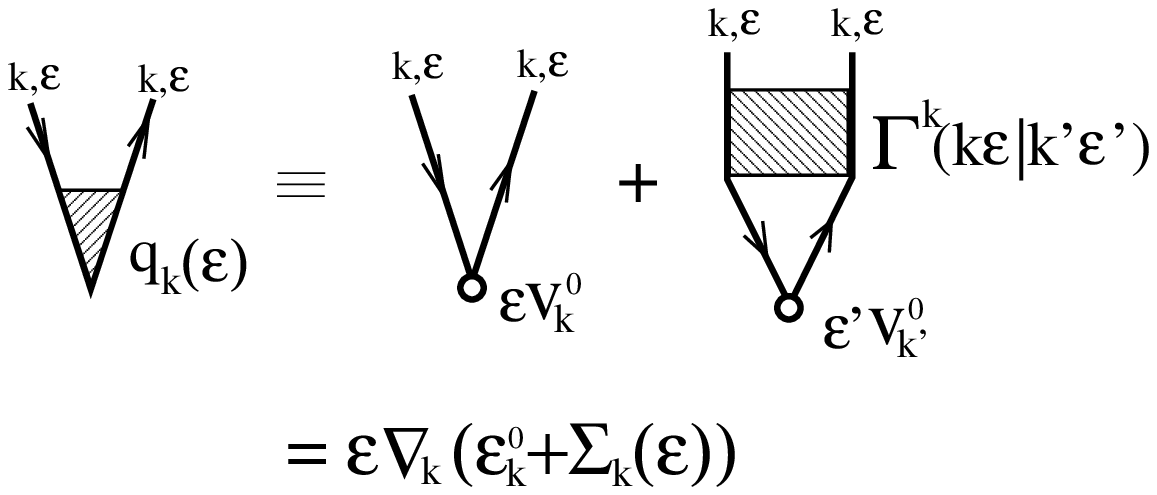,width=7cm}
\end{center}
\caption{
The Ward identity for the heat velocity ${\vec q}_\k(\e)$
derived in this section.
This identity assures that ${\vec q}_\k(\e=0)=0$.
}
  \label{fig:qi2}
\end{figure}

%When we put $p_\mu=0$ for $\mu\ne i$ ($i=$1,2 or 3)
%in eq.(\ref{eqn:Lqi}), then 
%%
%\begin{eqnarray}
%\Lambda_i^Q(k+p_i,k) 
% = \e \cdot \frac{\e_{{\bf k}+p_i}^0 + \Sigma(k+p_i) 
%   - \e_{\bf k}^0 - \Sigma(k)}{p_i}  .
%\end{eqnarray}
%
By putting $p_\mu=0$ for $\mu\ne i$ ($i=$1,2 or 3) 
in eq.(\ref{eqn:X4-2}) and taking the limit 
$p_i\rightarrow 0$, we obtain the $i$-component
of the heat velocity $q_i(k)$ as follows:
\begin{eqnarray}
q_i(k) &\equiv&
\lim_{p_i\rightarrow0} \Lambda_i^Q(k+p_i,k) 
 \nonumber \\
 &=& \e \frac{\d}{\d k_i}(\e_{\bf k}^0 + \Sigma(k))
 \label{eqn:qi}
\end{eqnarray}
for $i=1,2,3$, 
which is equivalent to the Ward identity
for the heat velocity, eq.(\ref{eqn:Ward}).
%According to eq.(\ref{eqn:qi}),
%the bare current (without VC's)
%is given by ${\vec q}^0(\k)= \e{\vec v}(\k)$.
By constructing the Bethe-Salpeter equation
%according to the microscopic Fermi liquid theory
 \cite{AGD,Nozieres},
${\vec q}(\k,\e)$ in eq.(\ref{eqn:qi}) is 
expressed by using the $k$-limit four-point vertex
$\Gamma^{k}(\k\e,\k'\e')$ as follows:
\begin{eqnarray}
 {\vec q}_\k(\e)
&=& \e {\vec \nabla}_k (\e_{\bf k}^0 + \Sigma(p)) 
 \nonumber \\
&=& \e {\vec v}_\k^0
 + \frac1{(2\pi)^3}\sum_{\k'} \int\frac{d\e'}{2\pi i}
  \Gamma^{k}(\k\e,\k'\e')
  \{ G_{\k'}(\e')^2 \}^k \cdot \e'{\vec v}_{\k'}^0
 \label{eqn:qi2}
\end{eqnarray}
in terms of the zero-temperature perturbation theory,
which is diagrammatically shown 
in fig.\ref{fig:qi2}.
In the finite temperature perturbation theory,
${\vec q}_\p(\e)$ is expressed as eq.(\ref{eqn:q}).
We stress that
eq.(\ref{eqn:qi2}) is satisfied only when
we take account of all the diagrams
for ${\Gamma}^k(k\e,k'\e')$ which are
given by the functional derivative
$\delta\Sigma/\delta G$.
(see Appendix D.)
%%%%%%%%%%%%%%%%%%%%%%%%%%%%%%%%%%%%%%%%%%%%%%%%%%%%%%
%%%%%%%%%%%%%%%%%%%%%%%%%%%%%%%%%%%%%%%%%%%%%%%%%%%%%%

Finally, 
we discuss the following two restrictions
assumed in the proof of the Ward identity:

(i) In the above discussion,
we assumed that the potential term in $h(z)$ is local,
which is not true if the range of the interaction is finite.
In this case,
a correction term $C(x,y;z)$ 
given in eq.(\ref{eqn:correction})
should be added to (\ref{eqn:X4}),
as discussed in Appendix C.
This correction term gives rise to
the additional heat velocity
$\Delta{\vec q}(k)$ given by eq.(\ref{eqn:Delta-q}).
Note that $\Delta{\vec q}_\k(\e)$
vanishes identically in the case of
the on-site Coulomb potential.
As a result,
the Ward identity for the heat velocity
in the case of general two-body interactions
is given by
\begin{eqnarray}
 {\vec q}(\k,\e)
= \e {\vec v}_\k(\e) + \Delta{\vec q}_\k(\e) .
 \label{eqn:qi2-final}
\end{eqnarray}
%
%under the limit $|\q|\rightarrow 0$.
Fortunately, $\Delta{\vec q}(k)$ does not 
contribute to the transport coefficients
as discussed in Appendix C.
As a result,
we can use the expression for ${\vec q}(k)$
in eq.(\ref{eqn:qi2})
for the purpose of 
calculating $S$, $\kappa$ and $\nu$,
even if the potential $U(\k)$ is momentum-dependent.

(ii) Here we treated the space variables like
${\bf x}$, ${\bf y}$ and ${\bf z}$
as continuous ones for the simplicity.
However, it is easy to perform the similar
analysis for the tight-binding model,
by replacing the derivative of $\bf x$ with 
the differentiation.
For example, the local energy conservation law
is expressed as 
$\frac{\d}{\d t}h(r_l) + (j(r_{l+1})-j(r_{l-1}))/2a=0$,
where $a$ is the lattice spacing.
We stress that
the Ward identity in 
eq.(\ref{eqn:qi}) is rigorous also 
in the case of the tight-binding model.
Note that we give the another proof 
for the Ward identity
based on the diagrammatic technique
in Appendix D,
which is valid for general tight-binding models.

We comment that Langer studied the Ward identity for the heat velocity
in ref.\cite{Langer}.
Unfortunately, because of a mistake, an extra factor 
$-{\vec v}_\k G_\k^{-1}(\w)$ should be subtracted from
the r.h.s. of his Ward identity (i.e., the heat velocity), 
eq.(3.31) in ref.\cite{Langer};
see eq.(\ref{eqn:qi2}) or (\ref{eqn:qi2-final}) 
in the present paper.
In fact,
the factor $\k\cdot(\k+\q)$ in eqs. (3.16) and (3.22)
(in eqs. (3.19) and (3.23)) of ref.\cite{Langer}
should be replaced with $\k^2$ (with $(\k+\q)^2$).
This failure, which is fortunately not serious in studying
the transport coefficients at lower temperatures, 
becomes manifest if one studies the Ward identity
in terms of the $x$-representation,
like in the present study.

%%%%%%%%%%%%%%%%%%%%%%%%%%%%%%%%%%%%%%%%
% formula for the Nernst coefficient
%%%%%%%%%%%%%%%%%%%%%%%%%%%%%%%%%%%%%%%%
\section{Formula for the Nernst coefficient}

According to the linear response theory,
the Nernst coefficient is given by
\begin{eqnarray}
\nu = \left[ \frac{-\a_{yx}}{\s} - S\tan\theta_{\rm H} \right]/B ,
 \label{eqn:nu}
\end{eqnarray}
where 
${\hat \a}({\bf B}) \equiv {\hat L}^{12}({\bf B})/T^2$
is called the Peltier tensor
(${\vec J}={\hat a}(-{\vec \nabla}T)$), and 
$\tan\theta_{\rm H}\equiv \s_{xy}/{\s}$ is the Hall angle.
Note that 
$\a_{yx}({\bf B}) = {L}_{yx}^{12}({\bf B})/T^2
 = {L}_{xy}^{21}(-{\bf B})/T^2 = -{L}_{xy}^{21}({\bf B})/T^2$
because of the Onsager relation.
In addition,
$\a_{yx}({\bf B}) = -\a_{xy}({\bf B})$
in the presence of the four-fold symmetry
along the magnetic filed ${\bf B}$.

In this section,
we investigate the off-diagonal Peltier coefficient $\a_{yx}$
due to the Lorentz force
to derive the expression for the Nernst coefficient.
Up to now,
the general expression for the Hall coefficient \cite{Kohno,Fukuyama}
and that for the magnetoresistance
 \cite{MR}
were derived by using the Fermi liquid theory
based on the Kubo formula.
These works enabled us to perform 
numerical calculations for the Hubbard model
 \cite{Kontani,MR-HTSC,S-HTSC}
within the conservation approximation
as Baym and Kadanoff
 \cite{Baym}.
Hereafter, we derive the general expression
for the Nernst coefficient by using the technique
developed in refs.
 \cite{Kohno}, \cite{MR} and \cite{Fukuyama}.

For the present purpose,
we have to include the external magnetic field.
In the presence of the vector potential, 
the hopping parameter $t_{ij}$ in eq.(\ref{eqn:H0})
is multiplied by the Peierls phase factor
 \cite{MR}:
\begin{eqnarray}
t_{ij} \rightarrow t_{ij} \exp[ ie({\bf A}_i+{\bf A}_j)
 \cdot({\bf r}_i-{\bf r}_j)/2 ] ,
 \label{eqn:Peierls}
\end{eqnarray}
where ${\bf A}_i$ is the external vector potential 
at ${\bf r}_i$, and $e (<0)$ is the charge of an electron,
Here we introduce ${\bf A}_i$ as
\begin{eqnarray}
{\bf A}_i = {\bf A} e^{i{\bf p}\cdot {\bf r}_i} ,
 \label{eqn:A-phase}
\end{eqnarray}
where ${\bf A}$ is a constant vector.
In this case, the magnetic field is given by 
${\bf B} = i{\bf p} \times {\bf A}$
in the uniform limit, i.e., $|{\bf p}| \ll 1$
 \cite{Kohno,MR,Fukuyama}.
Bearing eqs.(\ref{eqn:Peierls}) and (\ref{eqn:A-phase}) in mind,
the current operator defined by eq.(\ref{eqn:def-j})
and the Hamiltonian are given by
 \cite{MR}
\begin{eqnarray}
j_\nu^{B}(\p\!=\! 0) 
 &=&  j_\nu({\bf 0}) - eA_\a \cdot j_{\nu\a}(-\p) ,
 \label{eqn:expand1}
 \\
H_{B}&=& H_{B=0} - \ eA_\a j_\a(-\p) ,
 \label{eqn:expand2}
\end{eqnarray}
in the tight-binding model up to $O(A)$.
Here and hereafter,
the summation with respect to the suffix 
which appears twice is taken implicitly.
$j_\a(\p)$ is given in eq.(\ref{eqn:vero-q}),
and 
\begin{eqnarray}
j_{\a\b}(\p) = e\sum_\k (\d_\a \d_\b \e_\k^0)
 c_{\k-\p/2}^\dagger c_{\k+\p/2} .
\end{eqnarray}

To derive the Nernst coefficient,
we have to calculate the $L_{xy}^{21}$
under the magnetic field, which is given by
\begin{eqnarray}
L_{xy}^{21}({\p},i\w_l;{\bf A})
 = \frac{-T}{\w_l} \int_0^\beta d\tau e^{i\w_l\tau} \frac1Z
 {\rm Tr} \left\{ e^{-\beta H_{B}} T_\tau 
 {j_x^Q}^B({\p},\tau) j_y^B({\bf 0},0) \right\} ,
 \label{eqn:L21-qA}
\end{eqnarray}
and take derivative of eq.(\ref{eqn:L21-qA})
with respect to $q_\rho$ and $A_\s$
up to the first order.
$Z$ is the partition function.
By taking eqs.(\ref{eqn:expand1}) and (\ref{eqn:expand2})
into account, the $A_\s,q_\rho$-derivative of 
$L_{xy}^{12}({\p},i\w_l;{\bf A})$ is given by
%the $B$-linear term of $\a_{xy}(i\w_l)$ is given by
%
\begin{eqnarray}
-\a_{yx}(i\w_l)&\equiv& \sum_{\s\rho}C_{xy}^{\rho\s}(i\w_l)
 (ip_\rho \cdot A_\s) ,
 \\
C_{xy}^{\rho\s}(i\w_l)&\equiv&
 \left. \frac{-i}{T^2}\frac{\d^2}{\d p_\rho \d A_\sigma} 
 L_{xy}^{21}({\p},\w_l;{\bf A}) \right|_{{\bf p}={\bf A}=0} .
 \label{eqn:axy-iw}
\end{eqnarray}
Below, we see that the dc-Peltier coefficient $\a_{yx}$ is given by 
the analytic continuation of $\a_{yx}(i\w_l)$.

\begin{figure}
\begin{center}
\epsfig{file=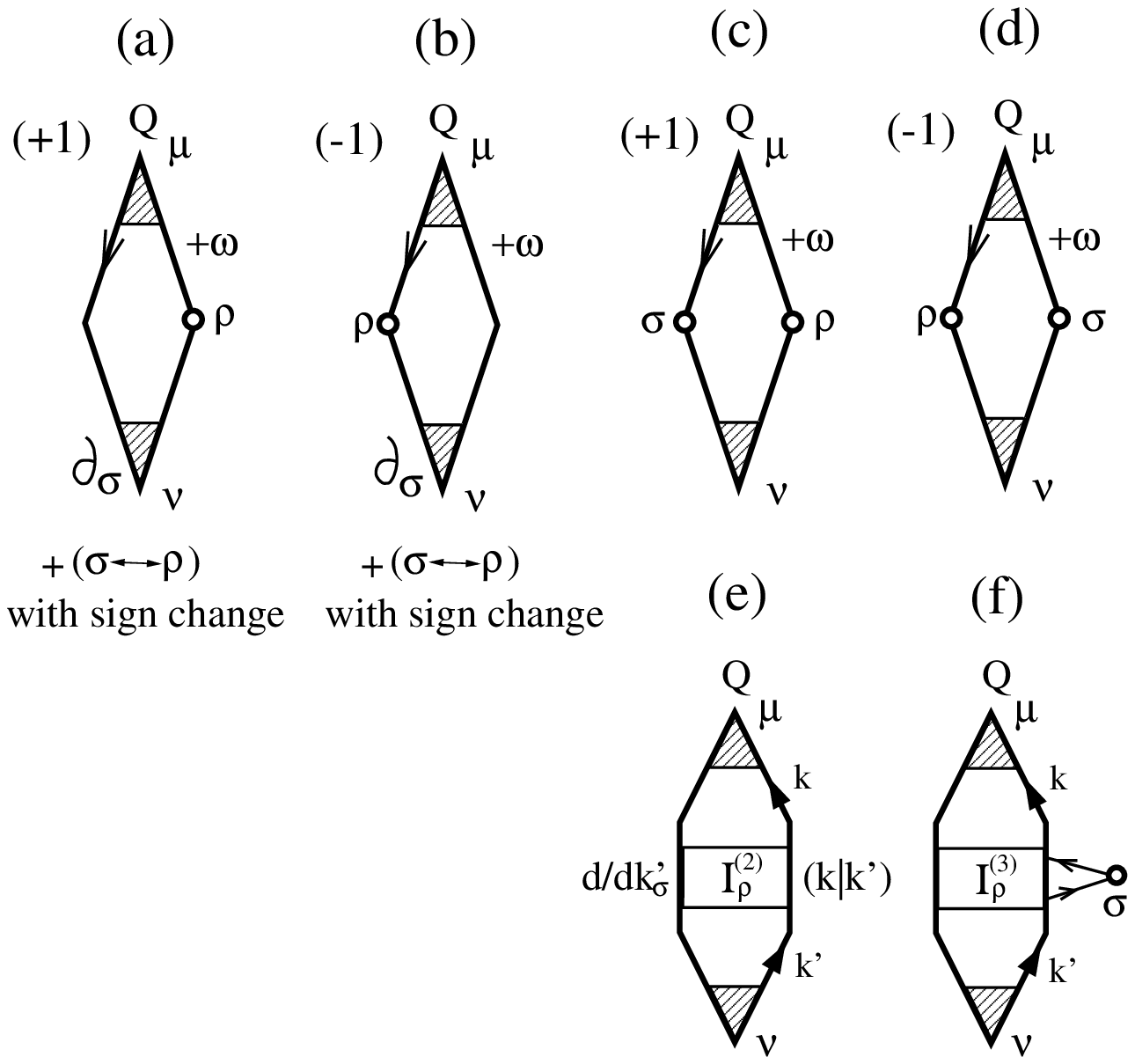,width=8cm}
\end{center}
\caption{}
\begin{center}
All the diagrams for $\a_{\mu\nu}$.
The symbol ``$\circ \rho$'' on each line
represents the momentum derivative, ``$\d_\rho$''.
The notations in the diagrams are explained
in ref.\cite{MR} in detail.
\end{center}
  \label{fig:Ap-N2}
\end{figure}

The diagrammatic expression for $C_{xy}^{\rho\s}(i\w_l)$
is very complicated, containing six-point vertices.
Fortunately,
as for the most divergent term with respect to $\gamma^{-1}$,
they can be collected into a small number of simpler diagrams
as shown in fig.\ref{fig:Ap-N2}
by taking the Ward identity into account
\cite{Kohno,MR,Fukuyama}.
We can perform the present calculation for $C_{xy}^{\rho\s}(i\w_l)$
in a similar way to that for $\s_{xy}$ in ref.
 \cite{Kohno},
only by replacing ${j}_{\mu=x}$ with ${j}^Q_{\mu=x}$ 
and using the Ward idntity for the heat velocity.
As a result, we obtain the following result:
\begin{eqnarray}
C_{xy}^{\rho\s}(i\w_l) &=&
 \frac{i e^2}{\w_l} \sum_{\k,i\e_n} {\Lambda}_{\k x}^Q(i\e_n;i\w_l)
 \left\{ [G\dalt_{\rho} G^+]\d_\s
   -[G\dalt_{\s} G^+]\d_\rho \right\}
  \Lambda_{\k y}(i\e_n;i\w_l)
 \nonumber \\
& &+ \frac{i e^2}{\w_l} \sum_{\k,i\e_n} {\Lambda}_{\k x}^Q(i\e_n;i\w_l)
 [\d_\rho G^+ \cdot \d_{\s} G
  -\d_{\s} G^+ \cdot \d_\rho G]
  \Lambda_{\k y}(i\e_n;i\w_l) ,
 \label{eqn:Cxy}
\end{eqnarray}
where 
$G^+ \equiv G_\k(\e_l+\w_l)$,
$G \equiv G_\k(\e_l)$, and 
$[A\dalt_\a B] \equiv A \!\cdot\! \d_\a B - B \!\cdot\! \d_\a A$.
${\Lambda}_{\k x}^Q(i\e_n;i\w_l)$
is the three point vertex for the heat current
given in eq.(\ref{eqn:L22-Matsubara}).
%which is derived in \S III.
Equation (\ref{eqn:Cxy}) is described in
(a)-(d) fig.\ref{fig:Ap-N2}.
Here, we neglect the diagrams
(e) and (f) because their contribution
is less singular with respect to $\gamma^{-1}$
 \cite{Kohno}.
%%%%%%%%%%%%%%%%%%%%%%%%%%%%%%%%%%%%%%%%%%%%%%%%%%%%%%
%%%%%%%%%%%%%%%%%%%%%%%%%%%%%%%%%%%%%%%%%%%%%%%%%%%%%%
Here, we assume that
the magnetic field ${\bf B}$ is parallel to the $z$-axis.
Then, we can easily check that 
$C_{xy}^{\rho\s}(i\w_l)$ in eq.(\ref{eqn:Cxy}) 
is expressed as 
$C_{xy}^{\rho=x,\s=y}(i\w_l)\cdot \e_{z\rho\s}$.
This fact assures that
$\a_{xy}(i\w_l)$ given in eq.(\ref{eqn:axy-iw})
is gauge-invariant, that is,
\begin{eqnarray}
-\a_{yx}(i\w_l)= C_{xy}^{\rho=x,\s=y}(i\w_l)\cdot B_z.
\end{eqnarray}
We note that eq.(\ref{eqn:Cxy}) is equivalent
to $\s_{xy}(i\w_l)/T$ if $\Lambda_{\k y}^Q$
is replaced with $\Lambda_{\k y}$;
see ``A'' and ``B'' in p.632 of ref.
 \cite{Kohno}.

In performing the analytic continuation of eq.(\ref{eqn:Cxy}),
the most divergent term with resect to $\gamma^{-1}$
is given by the replacements
$G(\e_n+\w_l) \rightarrow G^{\rm R}$ and
$G(\e_n) \rightarrow G^{\rm A}$.
Taking account of the relation
$\d_\rho G^{\rm R}(\e) 
 = G^{\rm R}(\e)^2 (v_{\k \rho}(\e)-i\d_\rho \gamma_\k(\e))$,
%$\d_\rho G^{\rm A}(\e) 
% = G^{\rm A}(\e)^2 (v_{\k \rho}(\e)+i\d_\rho \gamma_\k(\e))$,
the dc-Peltier coefficient $\a_{xy}(\w+i\delta)|_{\w=0}$
is obtained as
\begin{eqnarray}
-\a_{yx} &=& B \cdot \frac{e^2}{T}\sum_\k\int \frac{d\e}{\pi} 
 \left(-\frac{\d f}{\d\e}\right) 
 \bigg\{ |{\rm Im}G_\k(\e)||G_\k(\e)|^2 
  Q_{\k x} [v_{\k x} \d_y - v_{\k y} \d_x ] J_{\k y} 
 \nonumber \\
& & + |G_\k(\e)|^4 Q_{\k x} J_{\k y}
 [-v_{\k x} \d_y \gamma_{\k} + v_{\k y} \d_x \gamma_{\k}] 
 \bigg\} ,
 \label{eqn:axy1}
\end{eqnarray}
where ${\vec Q}_{\k}(\e)$ is the total heat current
introduced in eq.(\ref{eqn:Q})
We stress that ${\vec Q}_{\k}(\e=0)=0$ at $T=0$,
as is discussed in \S II.

It is instructive to make a comparison
between $\a_{xy}$ and $\s_{xy}/eT$:
The latter is given by 
eq.(\ref{eqn:axy1})
by replacing $Q_x$ with $J_x$.
%(In this case, eq.(\ref{eqn:axy2}) is valid
%even in the absence of the four-fold symmetry.)
In this case,
the second term of eq.(\ref{eqn:axy1}),
which contains the $k$-derivative of $\gamma_\k$,
vanishes identically
because of the Onsager relation
$\s_{xy}({\bf B})=-\s_{yx}({\bf B})$.
As a result,
the general expression for $\s_{xy}$
given in eq.(3.38) of ref.
 \cite{Kohno}
is reproduced.

If the system has the four-fold symmetry along the $z$-axis,
then $\a_{xy}({\bf B})=-\a_{yx}({\bf B})$.
In this case, considering that
$|{\rm Im}G_\k(\e)||G_\k(\e)|^2 = |G_\k(\e)|^4 \gamma_\k$,
eq.(\ref{eqn:axy1}) can be rewritten as
 \cite{Kontani,Kanki}
\begin{eqnarray}
-\a_{yx} = \a_{xy} &=& B \cdot \frac{e^2}{T}\sum_\k\int \frac{d\e}{2\pi} 
 \left(-\frac{\d f}{\d\e}\right) |{\rm Im}G_\k(\e)||G_\k(\e)|^2 
 \gamma_\k(\e) A_\k(\e),
 \label{eqn:axy2} \\
A_\k(\e)&=& \left( {\vec Q}_{\k}(\e) 
 \times ({\vec v}_\k(\e) \times {\vec \nabla})_z
 \left({\vec J}_{\k}(\e) / \gamma_\k(\e) \right)
 \right)_z 
 \nonumber \\
&=& |{\vec v}_{\k}(\e)|_{\perp} \left( {\vec Q}_{\k}(\e) 
 \times \frac{\d}{\d k_\parallel} 
 \left({\vec J}_{\k}(\e) / \gamma_\k(\e) \right)
 \right)_z ,
 \label{eqn:AAA} 
\end{eqnarray}
where $|{\vec v}_{\k}|_{\perp}= \sqrt{v_{\k x}^2 + v_{\k x}^2}$,
and $k_\parallel$ is the momentum on the $xy$-plane
along the Fermi surface, i.e.,
along the vector ${\vec e}_\parallel
 = ({\vec e}_z \times {\vec v}_\k)/|{\vec v}_{\k}|_{\perp}$.
As noted above,
eq.(\ref{eqn:axy2}) becomes $\s_{xy}/eT$
by replacing $Q_x$ with $J_x$;
see eq.(22) in ref. \cite{Kontani}.

It is notable that
$A_\k(\e)$ in eq.(\ref{eqn:AAA})
is rewritten as
\begin{eqnarray}
(\gamma_\k(\e)/|{\vec v}|_\perp) A_\k(\e)
 &=& ( Q_{\k x}J_{\k x} + Q_{\k y}J_{\k y} )
 \frac{\d \theta_\k^J}{\d k_\parallel}
 +  \left({\vec Q}_\k \times {\vec J}_\k\right)_z
 \frac{\d}{\d k_\parallel} \log(|{\vec J}_\k|/\gamma_\k) ,
 \label{eqn:gA}
\end{eqnarray}
where $\theta_\k^J = \tan^{-1}(J_{\k x}/J_{\k y})$.
In an interacting system without rotational symmetry,
the second term with $k_\parallel$-derivative of $\gamma_\k$ 
does not vanish in general 
since ${\vec J}$ is not parallel to ${\vec Q}$
owing to the VC's by ${\cal T}^{22}$.
In contrast,
${\vec v}_\k(\e)={\vec q}_\k(\e)/\e$ 
because of the Ward identity.
In ref.
 \cite{Nernst-HTSC},
based on the fluctuation-exchange (FLEX)+T-matrix approximation,
we studied the Nernst coefficient of the square lattice 
Hubbard model as an effective model for high-$T_c$ cuprates.
We found that the second term of eq.(\ref{eqn:gA})
gives the huge contribution in the pseudo-gap region
if the VC's for currents are taken into account
in a conserving way.
As a result, the origin of the abrupt increase of the 
Nernst coefficient under the pseudo-gap temperature 
is well understood.

%%%%%%%%%%%%%%%%%%%%%%%%%%%%%%%%%%%%%%%%
% Discussions
%%%%%%%%%%%%%%%%%%%%%%%%%%%%%%%%%%%%%%%%
\section{Discussions}
%%%%%%%%%%%%%%%%%%%%%%%%%%%%%%%%%%%%%%%%
% formula for the Nernst coefficient
%%%%%%%%%%%%%%%%%%%%%%%%%%%%%%%%%%%%%%%%
\subsection{Vertex Correction for Thermal Conductivity}

In previous sections,
we studied various analytical properties
for ${\vec q}_\k(\e)$ or ${\vec Q}_\k(\e)$,
using the Ward identity for the heat velocity
derived in \S III.
In this subsection,
we study a free dispersion model ($\e_\k= \k^2/2m$)
in the presence of the electron-electron interaction
without Umklapp processes.
This situation will be realized
in a tight-binding Hubbard model
when the density of carrier is low; $n\ll1$.
Here, we explicitly calculate 
the total heat current ${\vec Q}_\k(\e)$
in terms of the conserving approximation.
The present result explicitly shows that
${\vec J}_\k(\e) \ne {\vec Q}_\k(\e)/\e$.

Next, as a useful application of the expression 
for the transport coefficients derived in previous sections,
we study the thermal conductivity $\kappa$
in a free-dispersion model.
Because of the absence of Umklapp processes,
the ($T^2$-term of the)
resistivity $\rho$ of this system should be zero
even at finite temperatures.
In a microscopic study based on the Kubo formula,
this physical requirement is recovered
by taking account of
all the VC's for the current
given by the Ward identity
 \cite{Yamada}.
On the other hand,
%we show in the present section that
the thermal conductivity is finite
even in the absence of the Umklapp processes
because heat currents are not conserved
in the elastic normal scattering processes.
Hereafter,
%we study the role of the VC's
%for $\kappa$, and 
we derive the $T\gamma^{-1}$-linear term of $\kappa$
in the free dispersion model
in terms of the conserving approximation.
For this purpose, we can drop the second term 
of eq.(\ref{eqn:K-L22}) because
$L^{12}L^{21}/T^2L^{11} = T S^2\s \sim O(T^3\gamma^{-1})$.
The obtained result is exact within the second order 
perturbation with respect to $U$.

\begin{figure}
\begin{center}
\epsfig{file=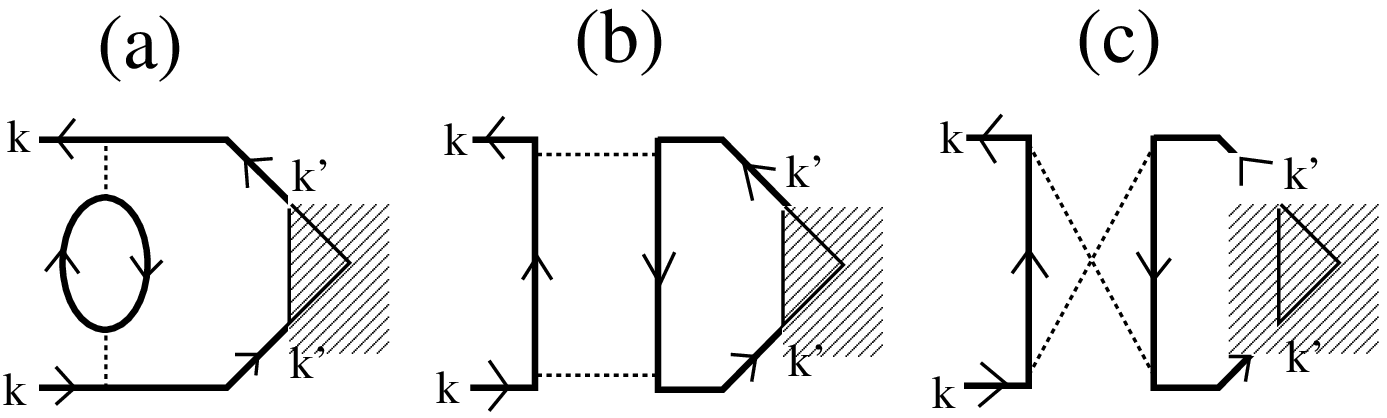,width=8cm}
\end{center}
\caption{
The vertex corrections (by ${\cal T}_{22}$)
for the heat/electron current
in the second order perturbation theory.
}
  \label{fig:QVC}
\end{figure}

First we consider the second order VC's
as shown in fig.\ref{fig:QVC}.
%Note that the result is easily generalized to 
%the exact expression beyond the second order.
Because ${\vec Q}_\k(\e=0)=0$,
we can write 
${\vec Q}_\k(\e) = {\vec C}_\k \cdot \e$
up to $O(\e)$.
The correction terms given by (a-c) in fig.\ref{fig:QVC}, 
$\Delta{\vec Q}_\k^{\mbox{(a-c)}}(\e)$,
are given by
\begin{eqnarray}
\Delta{\vec Q}_\k^{(r)}(\e)
 = U^2\sum_{\k'}\int \frac{d\e'}{4}
  \left[ {\rm cth}\frac{\e'-\e}{2T}-{\rm th}\frac{\e'}{2T} \right]
  {T}_{\k,\k'}^{(r)}(\e,\e')
  |G_{\k'}(\e')|^2 {\vec Q}_{\k'}(\e') ,
 \label{eqn:Qr}
\end{eqnarray}
where $r=a,b,c$. ${T}_{\k,\k'}^{(r)}(\e,\e')$
is a VC which is classified as ${\cal T}^{22}$:
Their functional form are given by 
\begin{eqnarray}
{T}_{\k,\k'}^{(a)}(\e,\e')
&=& \frac{2}{\pi} {\rm Im}\chi_{\k-\k'}^{0R}(\e-\e')
 \nonumber \\
&=& \sum_\p \int d\w 
 \left[-{\rm th}\frac{\w+\e}{2T}+{\rm th}\frac{\w+\e'}{2T} \right]
 \rho_{\k+\p}(\e+\w) \rho_{\k'+\p}(\e'+\w) ,
 \\
{T}_{\k,\k'}^{(b)}(\e,\e')
&=& {\cal T}_{\k,\k'}^{(a)}(\e,\e') ,
 \\
{T}_{\k,\k'}^{(c)}(\e,\e')
&=& \sum_\p \int d\w 
 \left[{\rm th}\frac{\w+\e}{2T}-{\rm th}\frac{\w-\e'}{2T} \right]
 \rho_{\k+\p}(\e+\w) \rho_{\k'+\p}(\e'-\w) , 
\end{eqnarray}
where $\rho_{\k}(\e)= \frac1\pi {\rm Im}G_\k(\e-i\delta)$
and $|G_\k(\e)|^2 = \pi\rho_\k(\e)/\gamma_\k(\e)$.
By expanding eq.(\ref{eqn:Qr}) with respect to
$\e$ and $T$ up to $O(\e^2,T^2)$ 
as was discussed in ref. \cite{Yamada}
and noticing that
$\displaystyle \left. \frac{\d}{\d\e} \int d\e' 
 [{\rm cth}\frac{\e'-\e}{2T}-{\rm th}\frac{\e'}{2T}]
 (\e'-\e)\e' \right|_{\e=0}= \frac13 (\pi T)^2$,
we obtain that
\begin{eqnarray}
\Delta{\vec Q}_\k^{(a)}
 &=& \frac{\e}{3} U^2\sum_{\k'\p}
 \pi \rho_{\k+\p}(0) \rho_{\k'+\p}(0) \rho_{\k'}(0)
 \frac{{\e}^2+(\pi T)^2}{2\gamma_{\k'}(\e)} {\vec C}_{\k'} ,
 \label{eqn:Qa} \\
\Delta{\vec Q}_\k^{(b)}
 &=& \Delta{\vec Q}_\k^{(a)} ,
 \label{eqn:Qb} \\
\Delta{\vec Q}_\k^{(c)}
 &=& -\Delta{\vec Q}_\k^{(a)} .
 \label{eqn:Qc} 
\end{eqnarray}
In deriving eq.(\ref{eqn:Qc}),
we have changed the integration variables
$(\k',\e') \rightarrow (-\k',-\e')$,
and used the relation
$\rho_{-\k}(0)=\rho_{\k}(0)$ and
${\vec Q}_{-\k}(-\e)={\vec Q}_{\k}(\e)$.
In general, within the FLEX approximation,
the Aslamazov-Larkin (AL) type VC's
by ${\cal T}^{22}$, which correspond to (b) and (c),
turn out to cancel out for the heat current.
%%%%%%%%%%%%%%%%%%%%%%%%%%%%%%%%%%%%%%%%%%%%%%%%%%%%%%
%%%%%%%%%%%%%%%%%%%%%%%%%%%%%%%%%%%%%%%%%%%%%%%%%%%%%%

\begin{figure}
\begin{center}
\epsfig{file=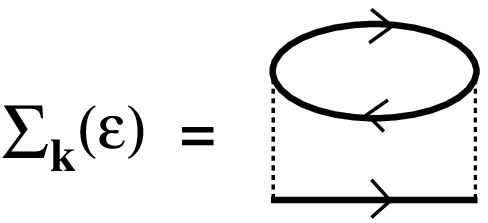,width=4cm}
\end{center}
\caption{
The self-energy given by the second order perturbation.
}
  \label{fig:self}
\end{figure}

In the same way,
the imaginary part of the self-energy, $\gamma_\k(\e)$,
is given by
\begin{eqnarray}
\gamma_\k(\e)
 = ({\e}^2+(\pi T)^2) \frac{U^2}{2} \sum_{\k'\p}
 \pi \rho_{\k+\p}(0) \rho_{\k'+\p}(0) \rho_{\k'}(0) ,
 \label{eqn:gamma}
\end{eqnarray}
which is shown in fig.\ref{fig:self}.
%%%%%%%%%%%%%%%%%%%%%%%%%%%%%%%%%%%%%%%%%%%%%%%%%%%%%%
%%%%%%%%%%%%%%%%%%%%%%%%%%%%%%%%%%%%%%%%%%%%%%%%%%%%%%

In a spherical system,
we can put ${\vec C}_\k = {C} \frac {\k'}{k_{\rm F}}$
on the Fermi surface.
Then, the total correction for ${\vec Q}$ is given by
\begin{eqnarray}
\Delta{\vec Q}_\k &\equiv& \sum_r^{a,b,c} \Delta{\vec Q}_\k^{(r)}
 \nonumber \\
 &=& 
 \frac{\e}{3Z} \sum_{\k'\p}
 \rho_{\k+\p}(0) \rho_{\k'+\p}(0) \rho_{\k'}(0)
  \cdot {C} \frac {\k'}{k_{\rm F}} ,
 \label{eqn:QQQ} \\
Z&=& \sum_{\k'\p}
 \rho_{\k+\p}(0) \rho_{\k'+\p}(0) \rho_{\k'}(0) .
 \label{eqn:Z} 
\end{eqnarray}
%
%where $|\k|=k_{\rm F}$.

Here, we put $\k= (0,0,k_{\rm F})$.
Then, the $z$-component of eq.(\ref{eqn:QQQ}) is given by 
\begin{eqnarray}
C\frac{\e}{3Z}
 \int dk' d\theta_{k'} d\phi_{k'} \int dp d\theta_{p} d\phi_{p}
 {k'}^2p^2 \sin\theta_{k'} \sin\theta_{p}\cdot 
 \delta(\e_{\k'}-\mu) \delta(\e_{\k+\p}-\mu) \delta(\e_{\k'+\p}-\mu)
 \cdot \cos\theta_{k'} .
 \label{eqn:int}
\end{eqnarray}
Note that in a free dispersion model,
$\rho_\k(0)= z\delta(z(\e_\k-\mu))= \delta(\e_\k-\mu)$,
where $\e_\k = \k^2/2m$ and 
$z$ is the renormalization factor.
%where we drop the renormalization factor $z$
%for the simplicity of the presentation.
By performing $k'$-integration,
$\theta_p$-integration, and 
$\phi_p,\phi_{k'}$-integrations successively,
eq.(\ref{eqn:int}) becomes
\begin{eqnarray}
& &C \frac{4\pi z^3 {m}^3 \e}{3Z}
\int d\theta_{k'} 
 \int_0^{2k_F^2\sin^2\theta_{k'}/(1-\cos\theta_{k'})} dp
 \frac{\sin\theta_{k'} \cos\theta_{k'}}
 {\sqrt{4k_F^2\sin^2\theta_{k'}-2p^2(1-\cos\theta_{k'})}}
 \nonumber \\
& &=
C \frac{4\pi {m}^3 \e}{3Z}
\int_0^\pi d\theta_{k'}
 \frac{\sin\theta_{k'} \cos\theta_{k'}}{\sqrt{2(1-\cos\theta_{k'})}}
 = C \frac{8\pi m^3 \e}{9Z} .
%= \frac{\e U^2}{3Z}\frac{8\pi m^3}{3} 
\end{eqnarray}
In the same way, $Z$ is calculated as
%{\vec J}_{\k}
\begin{eqnarray}
Z &=& 4\pi {m}^3 \int_0^\pi d\theta_{k'}
 \frac{\sin\theta_{k'}}{\sqrt{2(1-\cos\theta_{k'})}}
 \nonumber \\
&=& 8\pi m^3 .
\end{eqnarray}
As a result, $\Delta{\vec Q}_\k$ is given by
\begin{eqnarray}
\Delta{\vec Q}_\k = \frac\e9 C \frac {\k}{k_{\rm F}} 
 = \frac19 {\vec Q}_{\k} .
\end{eqnarray}

By solving the Bethe-Salpeter equation,
${\vec Q}= {\vec q}+ \Delta{\vec Q}$,
we get
\begin{eqnarray}
{\vec Q}_\k= \frac{9}{8} {\vec q}_\k ,
\end{eqnarray}
where ${\vec q}_\k = \e {\vec v}_\k$.
As a result, the thermal conductivity
within the second-order perturbation theory
is given by
\begin{eqnarray}   
\kappa = \frac{9}{8} \kappa^0 ,
 \\
\kappa^0 = \frac{\pi^2 n k_B^2 T}{6m \gamma} ,
 \label{eqn:kappa0}
\end{eqnarray}
where $\kappa^0$ is the result 
of the RTA,
where VC's are neglected.
Note that $m$ in eq.(\ref{eqn:kappa0}) is unrenormalized, 
and $n$ is the number of electrons in a unit volume.
Finally,
performing the momentum summations in eq.(\ref{eqn:gamma}),
$\gamma$ of order $U^2$ is given by
\begin{eqnarray}   
\gamma = (\e^2+(\pi T)^2) \frac{U^2m^3}{2\pi}. 
\end{eqnarray} 
In conclusion,
the vertex corrections slightly enhances
(by $\frac{9}{8}$ times)
the thermal conductivity 
in a three-dimensional free-dispersion model
within the second order perturbation theory.

It is instructive to 
make a comparison between
the role of VC's for the heat current 
and that for the electron current.
The VC's for the electron current
which correspond to fig.\ref{fig:QVC} (a)-(c)
are given by
\begin{eqnarray}
\Delta{\vec J}_\k^{(a)} &=&
  \Delta{\vec J}_\k^{(b)} = \Delta{\vec J}_\k^{(c)} 
 \nonumber \\
 &=& U^2\sum_{\k'\p}
 \pi \rho_{\k+\p}(0) \rho_{\k'+\p}(0) \rho_{\k'}(0)
 \frac{{\e}^2+(\pi T)^2}{2\gamma_{\k'}(\e)} {\vec J}_{\k'} ,
\end{eqnarray}
which was already derived in ref. \cite{Yamada}.
Here we put ${\vec J}_{\k}= D\frac{\k}{k_{\rm F}}$
on the Fermi surface.
Performing all the momentum integrations
in the spherical case as before,
we find that
\begin{eqnarray}
\Delta{\vec J}_\k &\equiv& \sum_r^{a,b,c} \Delta{\vec J}_\k^{(r)}
 \nonumber \\
 &=& 
 \frac3{Z} \sum_{\k'\p}
 \rho_{\k+\p}(0) \rho_{\k'+\p}(0) \rho_{\k'}(0) 
 \cdot D \frac{\k'}{k_F}
 \nonumber \\
 &=& D \frac{\k}{k_F} ,
\end{eqnarray}
where $Z$ is given in eq.(\ref{eqn:Z}).
As a result, the solution of the Bethe-Salpeter equation
${\vec J}= {\vec v}+ \Delta{\vec J}$ is given by
${\vec J}=\infty$,
which means that
the conductivity diverges in the absence of the Umklapp processes,
even at finite temperatures.
Thus, the important result in ref. \cite{Yamada}
is recovered.
On the other hand,
the thermal conductivity does not diverge
even in the absence of the Umklapp processes,
because the normal scattering process attenuates
the heat current.

%%%%%%%%%%%%%%%%%%%%%%%%%%%%%%%%%%%%%%%%%%%%%%%%%%%%%%%
\subsection{The TEP and the Nernst Coefficient}
%%%%%%%%%%%%%%%%%%%%%%%%%%%%%%%%%%%%%%%%%%%%%%%%%%%%%%%

In this section,
we discuss the effect of the anisotropy as well as
the role of the VC's for the TEP and the Nernst coefficient.
First, we discuss the validity of the Mott formula
for $S$ \cite{Mott}
which is given by 
\begin{eqnarray}
S&=& \frac{\pi^2k_{\rm B}^2T}{3e}
 \left[\frac{\d \ln \s(E)}{\d E}\right]_{E_{\rm F}} .
 \label{eqn:Sfree} 
\end{eqnarray}
It is easy to see that
eq.(\ref{eqn:Sfree}) is valid
even in the presence of Coulomb interactions,
if we define 
$\s(\e)\equiv e^2 \sum_\k |G_\k(\e)|^2 v_{\k x}(\e)J_{\k x}(\e)$
 \cite{Jonson}:
$\s$ and $S$ given by eqs.(\ref{eqn:L12}) and (\ref{eqn:sigma})
are rewritten using $\s(\e)$ as
\begin{eqnarray}
\s &=& \int \frac{d\e}{\pi} \left(-\frac{\d f}{\d \e}\right) \s(\e) ,
 \label{eqn:sig-MF}
 \\
S &=& \frac{1}{eT\s} 
  \int \frac{d\e}{\pi} \left(-\frac{\d f}{\d \e}\right) \e \s(\e) ,
 \label{eqn:S-MF}
\end{eqnarray}
At sufficiently lower temperatures,
eqs.(\ref{eqn:sig-MF}) and (\ref{eqn:S-MF}) become
\begin{eqnarray}
\s&=& \s(0) ,
 \\
S&=& \frac{\pi^2k_{\rm B}^2T}{3e\s} 
 \left. \frac{d\s(\e)}{d\e}\right|_{\e=0} .
\end{eqnarray}
As a result,
Mott formula is also satisfied in the case of 
electron-electron interaction.
%if we neglect the heat current carries by phonons.
Note that the renormalization factor $z$ does not appear
in eq.(\ref{eqn:Sfree}).

To analyze the TEP in more detail,
we rewrite the expression for $S$
by using the quasiparticle representation of the Green function,
eq.(\ref{eqn:Green-spect}),
which is possible at sufficiently low temperatures
in the Fermi liquid.
Using the relation
\begin{eqnarray}
\sum_\k &=& \int dS_k dk_\perp
 =  \int \frac{dS_k d\e_\k^0}{|v_\k^0|}
 \nonumber \\
 &=& \int \frac{dS_k d\e_\k^\ast}{z_\k|v_\k|} ,
\end{eqnarray}
where $S_k$ represents the Fermi surface
and ${k_\perp}$ is the momentum perpendicular 
to the Fermi surface,
we obtain the following expression: 
\begin{eqnarray}
S = \frac{e \pi^2k_{\rm B}^2 T}{3\s}\frac1{(2\pi)^3}
 \int \frac{dS_k}{z_\k|v_\k|} \frac{\d}{\d k_\perp}
 \left\{ \frac{v_{\k x}J_{\k x}}{|{\vec v}_\k|\gamma_\k}
 \right\}_{\e=\e_\k^\ast} ,
 \label{eqn:S-expand}
\end{eqnarray}
where we performed the $\e$-integration first
by assuming the relation $\gamma\ll T$.
In an anisotropic system, 
the $\k$-dependence of the integrand 
in eq.(\ref{eqn:S-expand}) may be strong.
In high-$T_{\rm c}$ cuprates,
for example, it is known that
the anisotropy of $\gamma_\k(0)$ on the Fermi surface
is very large because of the strong antiferromagnetic
fluctuations.
The point on the Fermi surface
where $\gamma_\k$ takes its minimum value
is called the ``cold spot'', and
the electrons around the cold spot
mainly contribute to the transport phenomena.
Because $\gamma_\k(\e_\k^\ast)$ has a huge
$k_\perp$-dependence in high-$T_{\rm c}$ cuprates
around the cold spot,
the sign of $S$ is almost determined by the sign of
$e\frac{\d \gamma_\k^{-1}(\e_\k^\ast)}{\d k_\perp}$
at the cold spot
 \cite{S-HTSC}.

Next, we discuss the Nernst coefficient.
Within the RTA,
the Nernst coefficient 
is derived from eqs.(\ref{eqn:axy1}), (\ref{eqn:L22})
and (\ref{eqn:sigma}) by dropping all the VC's
by ${\cal T}^{22}$.
In an isotropic system,
$\nu$ by RTA is expressed in a simple form as
 \cite{textbook1,textbook2}
\begin{eqnarray}
 \nu_{\rm RTA}
 &=& \frac{\pi^2k_{\rm B}^2T}{3m}
 \left[\frac{\d \tau(E)}{\d E}\right]_{E_{\rm F}} , 
 \label{eqn:Nfree}
\end{eqnarray}
where $\tau(\e)=1/2\gamma(\e)$ is the 
energy-dependent relaxation time
and $E_{\rm F}$ is the Fermi energy.
According to eq.(\ref{eqn:Nfree}),
$\nu$ is determined by the energy-dependence 
of the relaxation time.

Unfortunately,
eq. (\ref{eqn:Nfree}) will be too simple to analyze 
realistic metals with (strong) anisotropy.
For that purpose,
we perform the $\e$-integration of 
$\a_{xy}$ in eq.(\ref{eqn:axy2})
by using the quasiparticle representation.
The obtained expression for $\a_{xy}$ is given by
\begin{eqnarray}
\a_{xy} = B\frac{e^2 \pi^2k_{\rm B}^2 T}{12}\frac1{(2\pi)^3}
 \int \frac{dS_k}{z_\k|v_\k|}\frac{\d}{\d k_\perp}
  \left\{ 
 \left( {\vec Q'}_\k \times \frac{\d}{\d k_\parallel} 
 \left(\frac{{\vec J}_\k}{\gamma_\k} \right) \right)_z
 \frac{|{\vec v}_\k|_\perp}
 {|{\vec v}_\k|\gamma_\k}
 \right\}_{\e=\e_\k^\ast} ,
 \label{eqn:axy-expand}
\end{eqnarray}
where 
${\vec Q'}_\k(\e)\equiv {\vec Q}_\k(\e)/\e$
at zero temperature.
We stress that ${\vec Q'}_\k(\e=0)$ is finite
at $T=0$ as explained is \S II,
which leads to the relation $\nu \sim O(T\gamma^{-1})$.
We stress that ${\vec Q'}_\k(0)$ is not equal to 
${\vec J}_\k(0)$ in general,
because the VC's for heat current and the electron one
work in a different way;
see discussions in \S IV and \S V A.

We also comment that
the Mott formula type expression for $\a_{xy}$,
\begin{eqnarray}
 \a_{xy} \eqq \frac{\pi^2 k_{\rm B}^2 T}{3e}
 \left[\frac{\d \s_{xy}(E)}{\d E}\right]_{E_{\rm F}} 
 \label{eqn:Mott-a}
\end{eqnarray}
is obtained within the RTA,
by assuming that ${\vec Q}_\k(\e)=\e{\vec J}_\k(\e)$.
This assumption, however, will be totally violated
once we take the VC's into account.
As a result,
eq.(\ref{eqn:Mott-a}) is no more valid
in a correlated electron system.

Finally,
we discuss the Nernst coefficient in high-$T_c$ cuprates 
which increases drastically below the 
pseudo-gap temperature, $T^\ast$.
According to the numerical analysis
based on the conserving approximation
 \cite{Nernst-HTSC},
$k_\parallel$-dependence of $|{\vec J}_\k|$ becomes 
huge due to the VC caused by 
the strong superconducting fluctuations.
Moreover, ${\vec Q}_\k \times {\vec J}_\k$
is large because the VC is much effective
only for $ {\vec J}_\k$.
By considering eq.(\ref{eqn:gA}),
the growth of the Nernst coefficient
in high-$T_c$ cuprates under $T^\ast$
is caused by the enhancement of 
$\frac{\d}{\d k_\parallel}|{\vec J}_\k|$,
not by $\frac{\d}{\d E} \tau(E)$
 \cite{Nernst-HTSC}.

%%%%%%%%%%%%%%%%%%%%%%%%%%%%%%%%%%%%%%%%
% Summary
%%%%%%%%%%%%%%%%%%%%%%%%%%%%%%%%%%%%%%%%
\section{Summary}

In the present paper,
we have derived the general expressions
for $S$, $\kappa$ and $\nu$ 
in the presence of electron-electron interactions
based on the linear response theory 
for the thermoelectric transport phenomena.
%which had been developed by Luttinger or Mahan.
Each expression is ``exact''
as for the most divergent term 
with respect to $\gamma^{-1}$.
The heat velocity ${\vec q}_\k(\e)$,
which is required to calculate $S$, $\kappa$ and $\nu$,
is given by the Ward identity
with respect to the local energy conservation law.
We have studied the analytical properties of ${\vec q}_\k(\e)$ 
as well as the total heat current ${\vec Q}_\k(\e)$ in detail.

The expressions for $S$, $\kappa$ and $\nu$
derived in the present paper are summarized as follows.
Note that they are valid even if
the Coulomb potential $U(\k)$ has a momentum-dependence,
as discussed in Appendix C.
Here, $e (<0)$ is the charge of an electron.

(i) TEP: 
It is better to 
include the ``incoherent correction'',
which will be important in strongly correlated systems,
as is discussed in ref. \cite{S-BEDT}.
As a result, the final expression for $S$ is given by
\begin{eqnarray} 
 S = \frac{e}{T\s} 
 \sum_\k \int \frac{d\e}{\pi} \left(-\frac{\d f}{\d \e}\right)
 q_{\k x}(\e) \left[ \ |G_\k(\e)|^2 J_{\k x}(\e)
 - {\rm Re}\{G_\k^2(\e)\} v_{\k x}(\e) \ \right] ,
  \label{eqn:S-final}
\end{eqnarray}
where 
$\s=\s_{xx}$ is the electric conductivity,
${\vec v}_\k(\e)={\vec\nabla}_k(\e_\k^0+{\rm Re}\Sigma_\k(\e))$,
${\vec q}_\k(\e)= \e {\vec v}_\k(\e)$, and the total electron current
${\vec J}_{\k}(\e)$ is given in eq.(\ref{eqn:J}).

(ii) Thermal conductivity:
In the same way,
we include the incoherent correction.
Then, the final expression for $\kappa$ is given by
\begin{eqnarray} 
\kappa&=& \frac{1}{T} 
 \sum_\k \int \frac{d\e}{\pi} \left(-\frac{\d f}{\d \e}\right)
 q_{\k x}(\e) \left[ \ |G_\k(\e)|^2 Q_{\k x}(\e)
 - {\rm Re}\{G_\k^2(\e)\} q_{\k x}(\e) \ \right]
 \nonumber \\
& &- T S^2 \s ,
 \label{eqn:K-final}
\end{eqnarray}
where 
${\vec Q}_{\k}(\e)$ is given in eq.(\ref{eqn:Q}).
Note that ${\vec Q}_{\k}(\e)/\e \ne {\vec J}_{\k}(\e)$,
although the Ward identity 
${\vec q}_{\k}(\e)/\e = {\vec v}_{\k}(\e)$
is rigorously satisfied.

(iii) The expression for $\nu$ is given by
eq.(\ref{eqn:nu}), where $\a_{xy}$ is given by 
eq.(\ref{eqn:axy1}) or eq.(\ref{eqn:axy2}).
As for $\a_{xy}$ (and $\s_{xy}$),
no incoherent correction exists
as discussed in ref. \cite{S-BEDT}.

These derived expressions enable us to 
calculate the VC's in the framework of the 
conserving approximation.
In each expression, the factor 2 due to the spin degeneracy 
is taken into account.
%In this sense, they are useful for analysing
%strongly correlated electron systems.
We note that our expression are equivalent to
that of the relaxation time approximation (RTA),
if we drop all the vertex corrections in the formulae.
However, the RTA is dangerous because it may give unphysical results
owing to the lack of conservation laws.
In conclusion,
the present work gives us the fundamental framework for the
microscopic study of the thermoelectric transport phenomena
in strongly correlated electron systems.
Owing to the present work, the conserving approximation for 
thermoelectric transport coefficients becomes much practical
on the basis of the Fermi liquid theory.

%%%%%%%%%%%%%%%%%%%%%%%%%%%%%%%%%%%%%%%%%%%%%%%%%%%%%%%%%%%%%%
%%\acknowledgment
\vspace{5mm}
\begin{center}
{\bf acknowledgement}
\end{center}

The author is grateful to 
T. Saso, K. Yamada and K. Ueda 
for useful comments and discussions.
He is also grateful to one of the referees
for informing him of the existence of ref. \cite{Langer}.
%%%%%%%%%%%%%%%%%%%%%%%%%%%%%%%%%%%%%%%%%%%%%%%%%%%%%%%%%%%%%%

\appendix
%%%%%%%%%%%%%%%%%%%%%%%%%%%%%%%%%%%%%%%%
% Appendix A
%%%%%%%%%%%%%%%%%%%%%%%%%%%%%%%%%%%%%%%%
\def\Apeq{{\mbox{(\ref{eqn:jqM})}}}
\section{Another derivation of the heat current
operator, eq.$\Apeq$} 

In ref.
\cite{Jonson},
the authors derived the formula for $L^{12}(\w_l)$
under the condition that 
electron-phonon scattering and the impurity scattering
exist.
%The derived formula is same as eq.(\ref{eqn:L12M})
%in \S II within the most divergent term 
%with respect to $\gamma^{-1}$.
%In contrast, the present study is 
%based on the Ward identity for heat current
%which is valid for general kinds of
%electron-electron interaction as well as 
%electron-phonon one.
%As a result,
%we could confirm that
%eq.(\ref{eqn:L12M}) is valid for general two-body interactions.
In this appendix,
for an instructive purpose,
we derive eq.(\ref{eqn:L12M}) in \S II
in the case of the on-site Coulomb interaction
by using the similar technique used in ref.
\cite{Jonson}
This fact means that the heat current operator in the Hubbard model
can be rewritten as eq.(\ref{eqn:jqM}).

According to the equation of motion,
the following equations are satisfied:
\begin{eqnarray}
\frac{\d}{\d\tau}c_{k\s}^\dagger(\tau) 
&=& [H,c_{k\s}^\dagger(\tau)] \nonumber \\
&=& \e_k^0 c_{k\s}^\dagger + \frac U2 \sum_{K'q\s}
 c_{k-q,\s}^\dagger c_{k'+q/2,-\s}^\dagger c_{k'-q/2,-\s} ,
 \label{eqn:dtc1}
 \\
\frac{\d}{\d\tau}c_{k\s}(\tau) 
&=& [H,c_{k\s}(\tau)] \nonumber \\
&=& -\e_k^0 c_{k\s}^\dagger - \frac U2 \sum_{K'q\s}
 c_{k+q,\s} c_{k'+q/2,-\s}^\dagger c_{k'-q/2,-\s} .
 \label{eqn:dtc2}
\end{eqnarray}
Using $j_{\mu}^Q$ given in eq.(\ref{eqn:jq}) and
taking eqs. (\ref{eqn:dtc1}) and (\ref{eqn:dtc2}) into account,
we see that
\begin{eqnarray}
 \left\langle T_\tau j_{\mu}^Q(\tau) j_{\nu}(0)
  \right\rangle
&=&
\sum_{k\s} v_{k,\mu}^0 \e_k^0
  \left\langle T_\tau c_{k\s}^\dagger(\tau) c_{k\s}(\tau) j_{\nu}(0)
  \right\rangle
      \nonumber \\
& &+ \frac U2 \sum_{kk'q\s}\frac12 
  (v_{k+q/2,\mu}^0 + v_{k-q/2,\mu}^0)
  \left\langle T_\tau c_{k-q/2\s}^\dagger(\tau) c_{k+q/2\s}(\tau) 
  c_{k'+q/2,-\s}^\dagger(\tau) c_{k'-q/2,-\s}(\tau) j_{\nu}(0)
  \right\rangle
       \nonumber \\
&=&\lim_{\tau'\rightarrow\tau}
 \frac12 \left(\frac{\d}{\d\tau}-\frac{\d}{\d\tau'} \right)
\sum_{k\s} v_{k,\mu}^0
  \left\langle T_\tau c_{k\s}^\dagger(\tau) c_{k\s}(\tau') j_{\nu}(0)
  \right\rangle .
\end{eqnarray}
By inputting the above expression
in eq.(\ref{eqn:def-K}),
we can obtain the same expression as eq.(\ref{eqn:L12M}).
As a result, ${\vec j}^Q(\p=0,\w_l)$
can be expressed as eq.(\ref{eqn:jqM}).
We note that eq. (\ref{eqn:jqM}) is not exact
in the case of the finite range interactions.
Nonetheless, eq. (\ref{eqn:jqM}) is valid for the analysis
of the transport coefficient as for the most divergent term 
with respect to $\gamma^{-1}$, as discussed in \S III or
in Appendix C.
%Note that this proof is not general
%in that it depends on the specific form of the Hamiltonian.
%In addition, this proof would be complicated
%if we go beyond the on-site Coulomb interaction.

%%%%%%%%%%%%%%%%%%%%%%%%%%%%%%%%%%%%%%%%
% Appendix B
%%%%%%%%%%%%%%%%%%%%%%%%%%%%%%%%%%%%%%%%
\def\lm{{lm}}
\section{Definition of ${\cal T}^{\lm}(\p\e,\p'\e')$}
\begin{figure}
\begin{center}
\epsfig{file=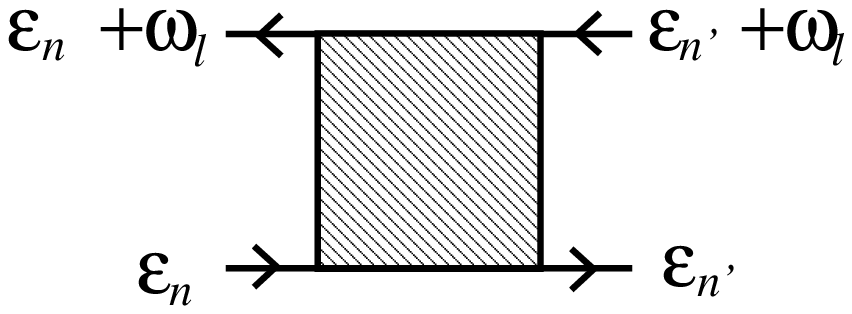,width=3.4cm}
\end{center}
\caption{
The diagrammatic expression for $\Gamma(i\e_n,i\e_{n'};i\w_l)$.
}
  \label{fig:VC_eew}
\end{figure}

\begin{figure}
\begin{center}
\epsfig{file=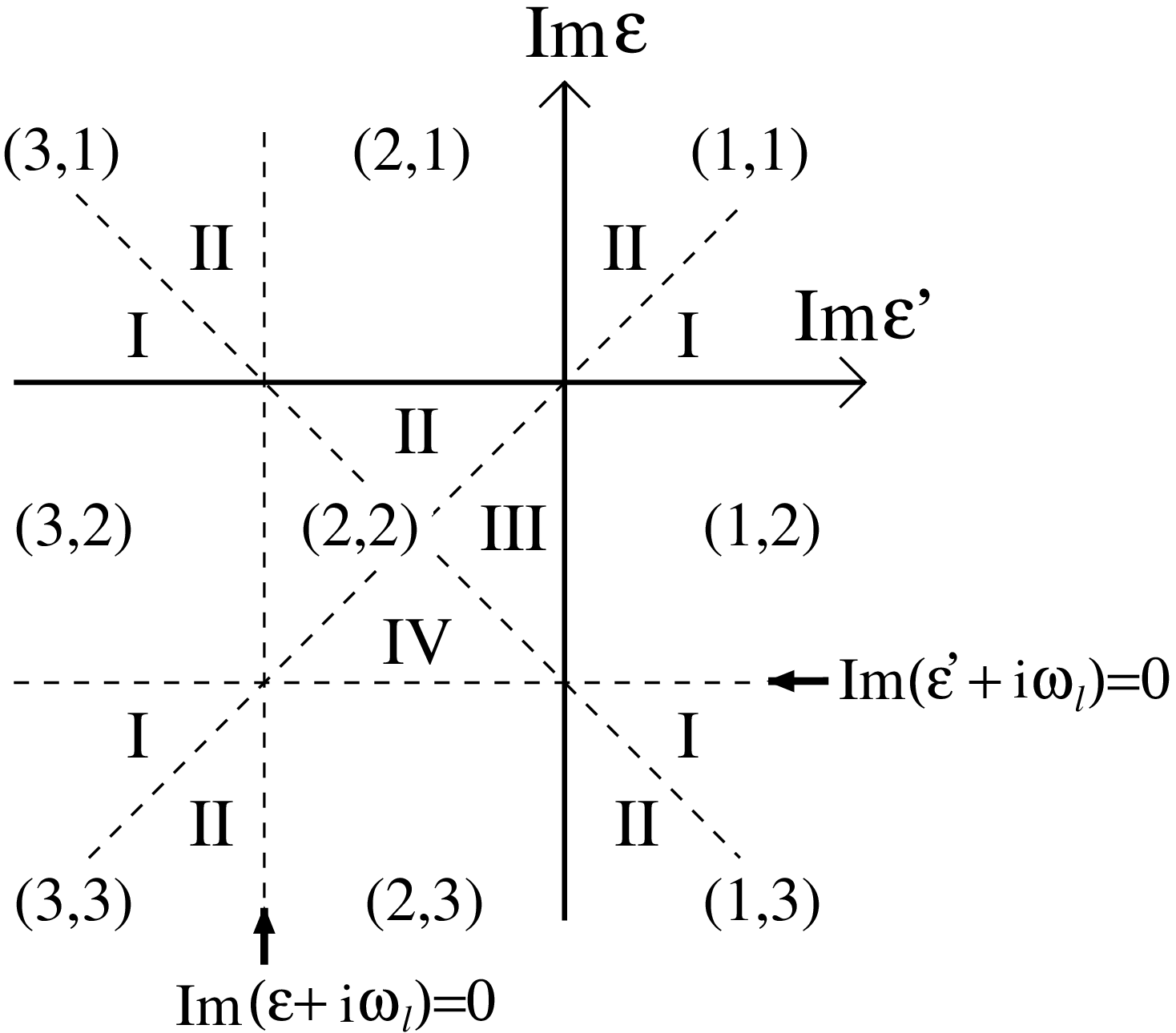,width=6cm}
\end{center}
\caption{
The definition of the region [$(l,m), N$].
}
  \label{fig:region_eew}
\end{figure}

Considering the convenience for readers, 
we list the expression for ${\cal T}^{\lm}(\p\e,\p'\e')$
introduced by Eliashberg in eq.(12) of 
ref.\cite{Eliashberg},
following the advice by referees.
Here we dropped the momentum suffixes for simplicity.
By taking the limit of $\w \rightarrow 0$,
they are given by
\begin{eqnarray}
{\cal T}^{11}(\e,\e')
 &=& {\rm th}\frac{\e'}{2T}\Gamma_{11}^\I
 + {\rm cth}\frac{\e'-\e}{2T}(\Gamma_{11}^\II-\Gamma_{11}^\I) ,
 \nonumber \\
{\cal T}^{12}(\e,\e') &=& 0 ,
 \nonumber \\
{\cal T}^{13}(\e,\e')
 &=& -{\rm th}\frac{\e'}{2T}\Gamma_{13}^\I
 - {\rm cth}\frac{\e'+\e}{2T}(\Gamma_{13}^\II-\Gamma_{13}^\I) ,
 \nonumber \\
{\cal T}^{21}(\e,\e')
 &=& {\rm th}\frac{\e'}{2T}\Gamma_{21} ,
 \nonumber \\
{\cal T}^{22}(\e,\e')
 &=& ({\rm cth}\frac{\e'-\e}{2T}-{\rm th}\frac{\e'}{2T})
  \Gamma_{22}^\II
 + ({\rm cth}\frac{\e'+\e}{2T}-{\rm cth}\frac{\e'-\e}{2T})
  \Gamma_{22}^\III
 + ({\rm th}\frac{\e'}{2T}-{\rm cth}\frac{\e'+\e}{2T})
  \Gamma_{22}^\IV ,
 \nonumber \\
{\cal T}^{23}(\e,\e')
 &=& -{\rm th}\frac{\e'}{2T}\Gamma_{23} ,
 \nonumber \\
{\cal T}^{31}(\e,\e')
 &=& {\rm th}\frac{\e'}{2T}\Gamma_{31}^\I
 + {\rm cth}\frac{\e'+\e}{2T}(\Gamma_{31}^\II-\Gamma_{31}^\I) ,
 \nonumber \\
{\cal T}^{32}(\e,\e') &=& 0 ,
 \nonumber \\
{\cal T}^{33}(\e,\e') 
 &=& -{\rm th}\frac{\e'}{2T}\Gamma_{33}^\I
 - {\rm cth}\frac{\e'-\e}{2T}(\Gamma_{33}^\II-\Gamma_{33}^\I) ,
\end{eqnarray}
where $\Gamma_{lm}^N \equiv \Gamma_{lm}^N(\e,\e')$ 
($l,m=1,2,3$, $N={\rm I,II,III,IV}$)
is a four-point vertex function,
which is introduced by the analytic continuation
of the four-point vertex function
$\Gamma(i\e_n,i\e_{n'};i\w_l)$ as shown in fig. \ref{fig:VC_eew}.
For instance,
$\Gamma_{11}^\I(\e,\e')$ comes from the analytic region 
[$(1,1), I$] in Fig. \ref{fig:region_eew}
(where $\w_l>0$)
and taking the limit $\w\rightarrow0$ at the final stage. 
There analytic properties are well studied in 
ref. \cite{Eliashberg}.

%%%%%%%%%%%%%%%%%%%%%%%%%%%%%%%%%%%%%%%%%%%%%%%%%%%%%%
%%%%%%%%%%%%%%%%%%%%%%%%%%%%%%%%%%%%%%%%%%%%%%%%%%%%%%
%%%%%%%%%%%%%%%%%%%%%%%%%%%%%%%%%%%%%%%%%%%%%%%%%%%%%%
%%%%%%%%%%%%%%%%%%%%%%%%%%%%%%%%%%%%%%%%%%%%%%%%%%%%%%

%%%%%%%%%%%%%%%%%%%%%%%%%%%%%%%%%%%%%%%%
% Appendix C
%%%%%%%%%%%%%%%%%%%%%%%%%%%%%%%%%%%%%%%%
\section{The Ward identity in the Case of the \\
Non-local Electron-Electron Interaction} 

In this appendix,
we show that
the expressions for $S$, $\kappa$ and $\nu$
derived in the present paper
is valid beyond the on-site Coulomb interaction.
For that purpose,
we reconsider the Ward identity
for the following Hamiltonian $H$
with a long-range interaction $U({\bf x}-{\bf y})$:
\begin{eqnarray}
h(z) &=& c^\dagger(z) 
 \left(\frac{-\hbar^2{\vec \nabla}^2}{2m}\right) c(z)
 + \frac12 c^\dagger(z)c(z)\int c^\dagger(r)c(r) V(z-r)d^4r ,
 \\
H&=& \int h(z)dz ,
\end{eqnarray}
where $V(x-y) \equiv U({\bf x}-{\bf y})\delta(x_0-y_0)$,
and $h(z)$ is the local Hamiltonian.
Hereafter, we drop the spin suffixes for simplicity.
In the same reason, we put $\mu=0$.

Here, it is easy to check that
\begin{eqnarray}
[h(z),c(x)]\delta(z_0-x_0) &=&
 \delta^4(x-z) \frac{\hbar^2{\vec \nabla}^2}{2m} c(x) 
 - \delta^4(z-x) \frac12 c(z) \int c^\dagger(r)c(r) V(z-r)d^4r
 \nonumber \\
& & -\frac12 c^\dagger(z)c(z)c^\dagger(z) V(z-x) ,
 \label{eqn:hz-c} 
\end{eqnarray}
\begin{eqnarray}
[H,c(x)] &=&
 \frac{\hbar^2{\vec \nabla}^2}{2m} c(x) 
 - \int c^\dagger(r)c(r) V(x-r)d^4r \cdot c(x) .
 \label{eqn:H-c}
\end{eqnarray}
By comparing eqs.(\ref{eqn:hz-c}) and (\ref{eqn:H-c})
and using the kinetic equation
$[H,c(x)]= -i\frac{\d}{\d x_0}c(x)$,
we obtain that
\begin{eqnarray}
[h(z),c(x)]\delta(z_0-x_0) &=&
 \delta^4(z-x)(-i)\frac{\d}{\d x_0}c(x)
 - \frac12 c^\dagger(z) c(z) c(x) V(z-x)
 \nonumber \\
& &+ \delta^4(z-x)\frac12 \int c^\dagger(r)c(r)V(r-x) d^4r 
  \cdot c(x) .
\end{eqnarray}
As a result,
\begin{eqnarray}
\langle T [h(z),c(x)]c^\dagger(y) \rangle \delta(x_0-z_0)
 &=& \delta^4(x-z) \frac{\d}{\d x_0} G(x-y) 
  - Y(x,y;z)V(x-z)
 \nonumber \\
 & & + \delta^4(x-z) \int Y(x,y;r)V(x-r)d^4r ,
\end{eqnarray}
where $T$ is a time-ordering operator, and
\begin{eqnarray}
Y(x,y;z) &\equiv& 
 \frac12 \langle T c^\dagger(z)c(z) c(x)c^\dagger(y) \rangle
  \nonumber \\
 &=& \frac12 \int\!\!\int G(x,x') \Lambda_0(x',y',z)G(y',y)d^4x'd^4y' ,
\end{eqnarray}
where $\Lambda_0(x',y',z)$ is the three-point vertex function
for the electron density; $\rho(z)=c^\dagger(z)c(z)$.
In the same way,
\begin{eqnarray}
\langle T c(x)[h(z),c^\dagger(y)] \rangle \delta(y_0-z_0)
 &=& \delta^4(y-z) \frac{\d}{\d y_0} G(x-y) 
  - Y(x,y;z)V(y-z)
 \nonumber \\
 & & + \delta^4(y-z) \int Y(x,y;r)V(y-r)d^4r .
\end{eqnarray}
As a result,
we find that the following correction term
\begin{eqnarray}
C(x,y;z)
&=& - Y(x,y;z)(V(x-z) + V(y-z))
 \nonumber \\
& &+ \int Y(x,y;r)(\delta^4(x-z)V(x-r) + \delta^4(y-z)V(y-r)) d^4r
 \label{eqn:correction}
\end{eqnarray}
is added to eq.(\ref{eqn:X4})
when $U(x-y)$ is a finite-range potential.
It is easy to see that
$C(x,y;z)=0$ if the potential is local
(i.e., $U(x-y)=U_0\delta({\bf x}-{\bf y})$).

%%%%%%%%%%%%%%%%%%%%%%%%%%%%%%%%%%%%%%%%%%%%%%%%%%%%%%
\begin{figure}
\begin{center}
\epsfig{file=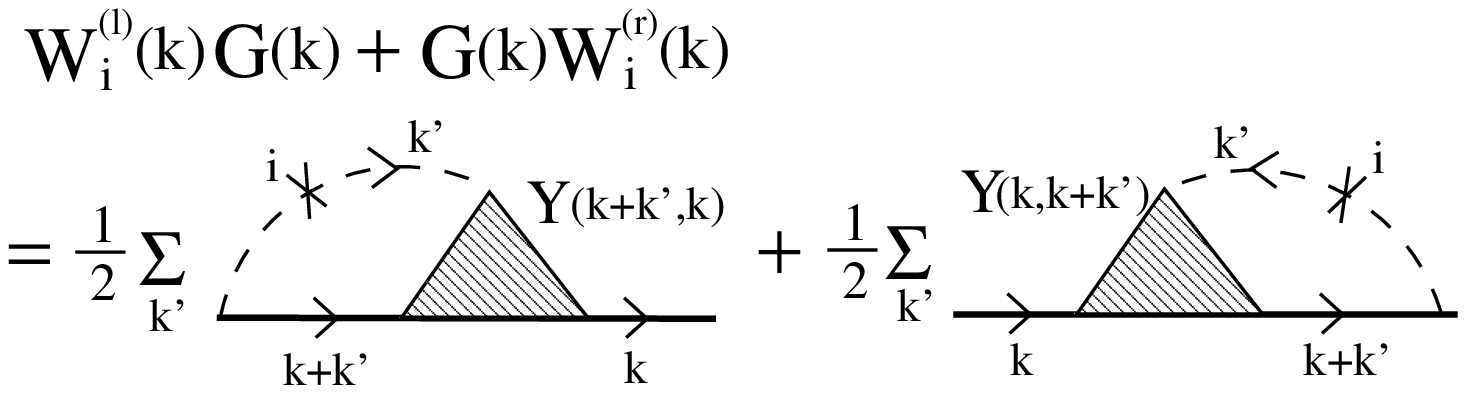,width=8cm}
\end{center}
\caption{}
\begin{center}
The diagrammatic expression for eq.(\ref{eqn:Wi}).
\end{center}
  \label{fig:XG}
\end{figure}
%%%%%%%%%%%%%%%%%%%%%%%%%%%%%%%%%%%%%%%%%%%%%%%%%%%%%%
%%%%%%%%%%%%%%%%%%%%%%%%%%%%%%%%%%%%%%%%%%%%%%%%%%%%%%
\begin{figure}
\begin{center}
\epsfig{file=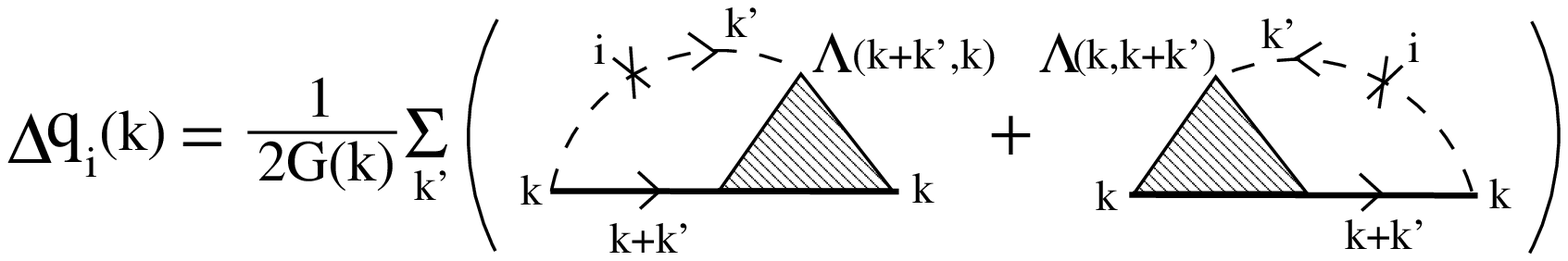,width=8cm}
\end{center}
\caption{}
\begin{center}
The diagrammatic expression for $\Delta{\vec q}(k)$
given by eq.(\ref{eqn:Delta-q}).
\end{center}
  \label{fig:WWW}
\end{figure}
%%%%%%%%%%%%%%%%%%%%%%%%%%%%%%%%%%%%%%%%%%%%%%%%%%%%%%
Now, we take the Fourier transformation of $C(x,y;z)$
according to eq.(\ref{eqn:Fourier}).
If we put $p_\mu=0$ for $\mu\ne i$ ($i$=1,2 or 3)
and $|p_i|\ll 1$,
$C(k;k+p_i)$ is given by 
\begin{eqnarray}
C(k;k+p_i) 
&\equiv& \int d^4xd^4yd^4z 
 \ C(x,y;z) e^{ik(x-y)+ip(x-z)}
 \nonumber \\
&=& ip_i \int d^4xd^4yd^4z 
 \ Y(x,y;z) \left[ V(x-z)\cdot(x_i - z_i) 
  + V(y-z)\cdot(y_i - z_i) \right]e^{ik(x-y)}
 \nonumber \\
& &+O(p_i^2) .
\end{eqnarray}
Because
the Fourier transformation of $V(x)\cdot x_i$
is given by $-i\frac{\d  U(k)}{\d k_i}$,
\begin{eqnarray}
\lim_{p_i\rightarrow0}\frac{C(k;k+p_i)}{ip_i}
&=& \int d^4k' \left( Y(k+k';k) + Y(k;k+k') \right)
 (-i)\frac{\d}{\d k_i'} U(k')
  \nonumber \\
&\equiv& \frac12 \left( W_i^{(l)}(k)G(k) + G(k)W_i^{(r)}(k) 
  \right) ,
 \label{eqn:Wi}
\end{eqnarray}
which is diagrammatically shown in fig. \ref{fig:XG}.
$W_i^{(l)}(k)$ and $W_i^{(r)}(k)$ are
introduced in the last line of eq.(\ref{eqn:Wi}).

Here we note that the energy dependence 
of $W_i^{(l,r)}(k)$ around the Fermi level
is same as that of $\Sigma(k)$, that is,
${\rm Re}W_i(k) \sim {\rm const.}$ and
${\rm Im}W_i(k) \approx k_0^2$
for $|k_0|\ll 1$.
This fact is easily recognized because
\begin{eqnarray}
\int d^4k' \left( Y(k+k';k) + Y(k;k+k') \right)
 i U(k')
 = \frac12 \left( \Sigma(k)G(k) + G(k)\Sigma(k) \right) ,
 \label{eqn:GS}
\end{eqnarray}
which is same as eq.(\ref{eqn:Wi})
except for the momentum derivative on $U$.

$W_i^{(l,r)}$ given in eq.(\ref{eqn:Wi})
provides the correction for the heat velocity
due to the non-locality of $h(z)$,
which we denote as $\Delta{\vec q}(k)$.
As shown in eq.(\ref{eqn:L12}) or eq.(\ref{eqn:L22}),
in the most divergent term for $L^{ij}$,
the (heat) current is connected with
$g^{(2)}(k) \equiv |G(k)|^2$ 
after the analytic continuation.
Bearing this fact in mind
and using ${\cal T}^{2i}= ({\cal T}^{1i}+{\cal T}^{3i})/2$
under the limit of $T\rightarrow T$
as discussed in ref. \cite{Eliashberg},
$\Delta {\vec q}(k)$ is given by
\begin{eqnarray}
\Delta {\vec q}(k) 
&=& \frac1{2G^{\rm R}G^{\rm A}}
 \left( \{{\rm Re}W_i^{(l)}\} G^{\rm R} 
 + G^{\rm A} \{{\rm Re}W_i^{(r)}\} \right)
 \nonumber \\
&=& {\rm Re}G^{-1}(k) \cdot {\rm Re}W_i(k) ,
 \label{eqn:Delta-q}
\end{eqnarray}
which should be added to ${\vec q}(k)$ given by 
eq.(\ref{eqn:qi2}) in \S III.
$W_i(k)$ is given by
\begin{eqnarray}
W_i(k) 
&=& W_i^{(l)}(k) = W_i^{(r)}(k)
%&\equiv& \frac12 (W_i^{(l)}(k)+W_i^{(r)}(k))
 \nonumber \\
&=& \frac12\int d^4k' 
 \left( \Lambda_0(k+k';k) + \Lambda_0(k;k+k') \right)
 G(k+k')(-i)\frac{\d}{\d k_i'} U(k') ,
\end{eqnarray}
which is expressed in fig.\ref{fig:WWW}.
In conclusion,
the Ward identity for the heat velocity
for general two-body interaction is given by
${\vec q}_\k(\e) = \e{\vec v}_\k(\e) + \Delta{\vec q}(k)$,
instead of eq.(\ref{eqn:qi2}).

Finally,
we study the contribution of $\Delta {\vec q}(k)$
to the transport coefficients. 
Let us assume that
$\gamma \ll T$ at sufficiently lower temperatures.
In this case,
the correction term for the TEP
due to $\Delta {\vec q}(k)$, $\Delta S$,
is calculated as
\begin{eqnarray}
\Delta S &\propto& \frac1{T\s} \sum_\k \int \frac{d\e}{\pi}
 \left(-\frac{\d f}{\d \e}\right) |G(k)|^2
 \Delta q_x(k) J_x(k)
  \nonumber \\
&=& \frac1{T\s} \sum_\k \left(-\frac{\d f}{\d \e}\right)_{\e_\k^\ast}
 \frac{z_\k \Delta q_x(\k,\e_\k^\ast) J_x(\k,\e_\k^\ast) }
 {\gamma_\k(\e_\k^\ast)} ,
 \label{eqn:DS}
\end{eqnarray}
where $\e_\k^\ast$ is the quasiparticle spectrum
given by the solution of 
${\rm Re}G^{-1}(\k,\e_\k^\ast)=0$.
Considering that $\Delta q_x(\k,\e_\k^\ast)=0$
because of eq.(\ref{eqn:Delta-q}),
we recognize that $\Delta S$ given by eq.(\ref{eqn:DS})
is zero.

In summary,
when $U(\k)$ is momentum-dependent, 
the corrections term for the heat velocity
$\Delta {\vec q}(k)$, given by eq.(\ref{eqn:Delta-q}),
emerges.
Fortunately,
its contribution to transport coefficients
would be negligible
when the concept of the quasiparticle is meaningful,
except for very high temperatures.
In conclusion,
the derived expressions for $S$, $\kappa$ and $\nu$,
given by eqs.(\ref{eqn:S-final}), (\ref{eqn:K-final})
and (\ref{eqn:axy2}) respectively,
are valid for general electron-electron interactions,
with the use of the heat velocity ${\vec q}(k)$ in eq.(\ref{eqn:qi2}).

%%%%%%%%%%%%%%%%%%%%%%%%%%%%%%%%%%%%%%%%
% Appendix D
%%%%%%%%%%%%%%%%%%%%%%%%%%%%%%%%%%%%%%%%
\section{Another Proof of the Ward Identity: \\
Based on the diagrammatic technique}

\begin{figure}
\begin{center}
\epsfig{file=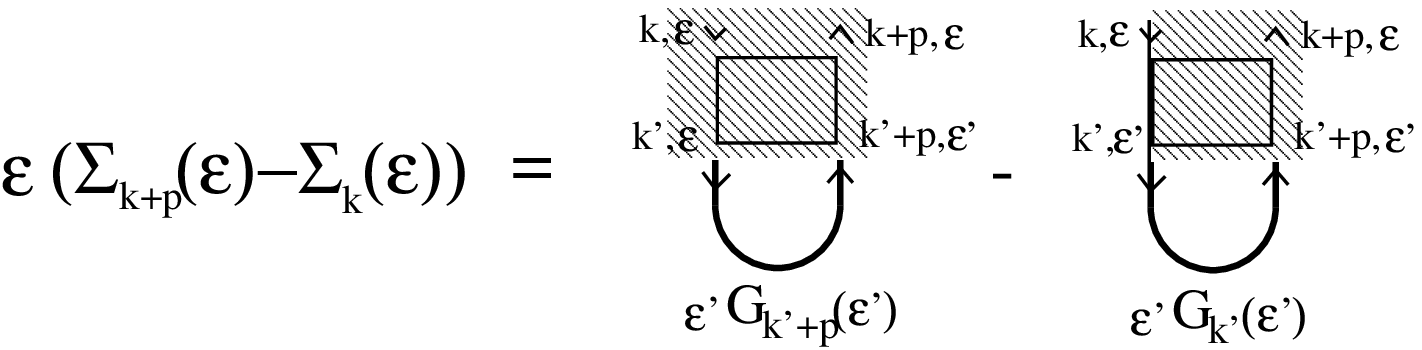,width=8cm}
\end{center}
\caption{
The generalized Ward identity 
with respect to the heat velocity.
}
  \label{fig:Ap-Ward}
\end{figure}

In the present Appendix, we give another proof that
the following generalized Ward identity is correct
in a tight-binding model
with on-site Coulomb interaction:
\begin{eqnarray}
& &\e[\Sigma_{\k+\p}(\e)-\Sigma_{\k}(\e)]
 \nonumber \\
& & \ \ \ =
 T\sum_{\e'\k'}\Gamma^I(\k\e;\k+\p,\e|\k'+\p,\e;\k'\e')
  [G_{\k'+\p}(\e') - G_{\k'}(\e')]\e' ,
 \label{eqn:Ap-GWI}
\end{eqnarray}
where $\Gamma^I$ is irreducible with respect to 
a particle-hole channel.
Equation (\ref{eqn:Ap-GWI})
is shown diagrammatically in fig.\ref{fig:Ap-Ward}.

%%%%%%%%%%%%%%%%%%%%%%%%%%%%%%%%%%%%%%%%%%%%%%%%%%%%%%
%%%%%%%%%%%%%%%%%%%%%%%%%%%%%%%%%%%%%%%%%%%%%%%%%%%%%%

Hereafter, we write that
$k=(\e,\k,\s)$. 
The $n$-th order skeleton diagrams for the self-energy 
are given by
\begin{eqnarray}
\Sigma_k^{(n)} &=& \sum_{P,\{k_i\}}\frac{A(P)}{n!}
 G_{k_1} G_{k_2} \cdots G_{k_{2n-1}}
 \nonumber \\
& & \times U(k,k_{\a_1}|k_{\a_2},k_{\a_3}) 
 U(k_{\a_4},k_{\a_5}|k_{\a_6},k_{\a_7}) \cdots
 U(k_{\a_{4n-4}},k_{\a_{4n-3}}|k_{\a_{4n-2}},k) ,
 \label{eqn:Sigma-n}
\end{eqnarray}
where $P$ represents the permutation
of $(4n-2)$-numbers, 
$(\a_1,\a_2,\cdots,\a_{4n-2}) = P(1,1,2,2,\cdots, 2n-1,2n-1)$.
$A(P)=\pm 1$ for a skelton diagram, 
and $A(P)=0$ for others.
$U(k_1,k_2|k_3,k_4)$ is the two-body interaction 
where $k_1,k_3$ are incoming and  $k_2,k_3$ are
outgoing, respectively
 \cite{Nozieres,Kuroda}.
%In the case of eq.(\ref{eqn:Hint}),
%$U(k_1,k_2|k_3,k_4)
% = U_{\s_1,\s_3}(\k_1-\k_2)\delta_{\k_1+\k_3,\k_2+\k_4}
% \delta_{\e_1+\e_3,\e_2+\e_4}
% \delta_{\s_1,\s_2}\delta_{\s_3,\s_4}$.
Here we consider the on-site Coulomb interaction:
$U(k_1,k_2|k_3,k_4)
 = U \delta_{\k_1+\k_3,\k_2+\k_4}
 \delta_{\e_1+\e_3,\e_2+\e_4}
 \delta_{\s_1,\s_2}\delta_{\s_3,\s_4} \delta_{\a_1,-\s_3}$.
Note that in eq.(\ref{eqn:Sigma-n}), 
the tadpole-type (Hartree-type) diagrams are dropped 
because they are $k$-independent.

\begin{figure}
\begin{center}
\epsfig{file=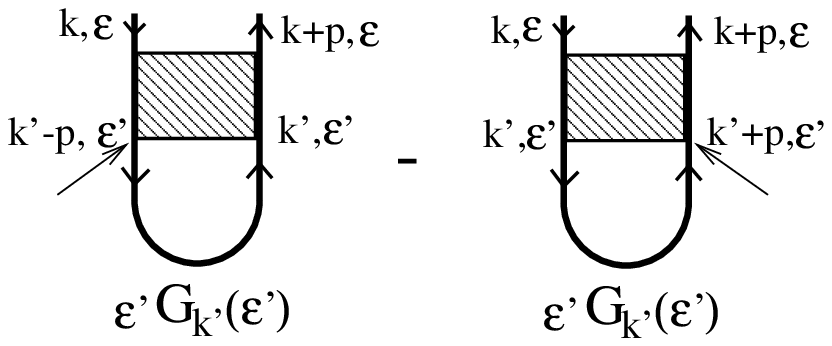,width=5cm}
\end{center}
\caption{}
%\begin{center}
The diagrammatic expression for eq.(\ref{eqn:Ap-GWI2}).
Here, the momentum conservation is violated by $\p$
only at the junction pointed by the arrow.
%\end{center}
  \label{fig:Ap-Ward2}
\end{figure}

Next, we consider the right-hand-site of 
eq.(\ref{eqn:Ap-GWI}),
which can be rewritten as
\begin{eqnarray}
 T\sum_{\e'\k'} 
 \left[ \Gamma^I(k;k+\p|k',k'-\p) -\Gamma^I(k;k+\p|k'+\p,k') \right]
 G_{\k'}(\e')\e' ,
 \label{eqn:Ap-GWI2}
\end{eqnarray}
which is shown in fig.\ref{fig:Ap-Ward2}.
%%%%%%%%%%%%%%%%%%%%%%%%%%%%%%%%%%%%%%%%%%%%%%%%%%%%%%
%%%%%%%%%%%%%%%%%%%%%%%%%%%%%%%%%%%%%%%%%%%%%%%%%%%%%%

Then, the $n$-th order skeleton diagrams 
for eq.(\ref{eqn:Ap-GWI2}) is expressed as
\begin{eqnarray}
& &\sum_{P,\{k_i\}}\frac{A(P)}{n!}
  G_{k_1} G_{k_2} \cdots G_{k_{2n-1}}
 \nonumber \\
& & \ \times \biggl\{ \ 
\Bigl[  \e_{\a_1} U(k,k_{\a_1}-\p|k_{\a_2},k_{\a_3})
  +\e_{\a_1} U(k,k_{\a_1}|k_{\a_2}k_{\a_3}-\p)
  -\e_{\a_2} U(k,k_{\a_1}|k_{\a_2}+\p,k_{\a_3}) \Bigr]
 \nonumber \\
& & \ \ \ \ \ \ \ \ \ \ \times
 U(k_{\a_{4}},k_{\a_{5}}|k_{\a_{6}},k_{\a_7}) \cdots 
 U(k_{\a_{4n-4}},k_{\a_{4n-3}}|k_{\a_{4n-2}},k+\p)
 \nonumber \\
& & \ \ \ \ \ \ +
\Bigl[  \e_{\a_5} U(k_{\a_{4}},k_{\a_{5}}-\p|k_{\a_{6}},k_{\a_7})
  +\e_{\a_7} U(k_{\a_{4}},k_{\a_{5}}|k_{\a_{6}},k_{\a_7}-\p)
 \nonumber \\
& & \ \ \ \ \ \ \ \ \ \ \ \
  -\e_{\a_4} U(k_{\a_{4}}+\p,k_{\a_{5}}|k_{\a_{6}},k_{\a_7})
  -\e_{\a_6} U(k_{\a_{4}},k_{\a_{5}}|k_{\a_{6}}+\p,k_{\a_7}) \Bigr]
 \nonumber \\
& & \ \ \ \ \ \ \ \ \ \ \times
 U(k,k_{\a_1}|k_{\a_2},k_{\a_3}) \cdots 
 U(k_{\a_{4n-4}},k_{\a_{4n-3}}|k_{\a_{4n-2}},k+\p)
 \nonumber \\
& & \ \ \ \ \ \ + \cdots
  \nonumber \\
& & \ \ \ \ \ \ +
\Bigl[  \e_{\a_{4n-3}} U(k_{\a_{4n-4}},k_{\a_{4n-3}}-\p|k_{\a_{4n-2}},k+\p)
  -\e_{\a_{4n-4}} U(k_{\a_{4n-4}}+\p,k_{\a_{4n-3}}|k_{\a_{4n-2}},k+\p)
 \nonumber \\
& & \ \ \ \ \ \ \ \ \ \ \ \ \ \ \ 
  -\e_{\a_{4n-2}} U(k_{\a_{4n-4}},k_{\a_{4n-3}}|k_{\a_{4n-2}}+\p,k+\p) \Bigr]
 \nonumber \\
& & \ \ \ \ \ \ \ \ \ \ \times
 U(k,k_{\a_1}|k_{\a_2},k_{\a_3}) \cdots 
 U(k_{\a_{4n-8}},k_{\a_{4n-7}}|k_{\a_{4n-6}},k_{\a_{4n-5}})
\ \biggr\} .
 \label{eqn:Ap-expand}
\end{eqnarray}

Because of the relation 
$U(k_1+\p,k_2|k_3,k_4)=U(k_1,k_2|k_3+\p,k_4)
 = U(k_1,k_2-\p|k_3,k_4)=U(k_1,k_2|k_3,k_4-\p)$
and the energy conservation
with respect to $U$,
we see that
eq.(\ref{eqn:Ap-expand}) $+ \ \e(\Sigma_{k}-\Sigma_{k+\p})=0$.
As a result,
the generalized Ward identity, eq.(\ref{eqn:Ap-GWI})
is proved in the framework of the microscopic perturbation theory.

By taking the limit $|\p|\rightarrow 0$
of eq.(\ref{eqn:Ap-GWI}),
we obtain
\begin{eqnarray}
\e_n \cdot {\vec \nabla}_k \Sigma_{\k}(\e_n)
 &=& T\sum_{\e_{n'}\k'} \lim_{\p\rightarrow 0}
 \Gamma^I(\k\e_n;\k+\p,\e_n|\k'+\p,\e_{n'};\k'\e_{n'}) \cdot
 G_{\k'+\p}(\e_{n'}) G_{\k'}(\e_{n'}) 
 \nonumber \\
& & \times
 \e_{n'} \left({\vec v}_{\k'}^0 
 + {\vec \nabla}_{k'} \Sigma_{\k'}(\e_{n'}) \right) .
 \label{eqn:Ap-GWI3-2}
\end{eqnarray}
Equation (\ref{eqn:Ap-GWI3-2}) is rewritten
by using the reducible four-point vertex $\Gamma$ as
\begin{eqnarray}
& &\e_n \cdot {\vec \nabla}_k \Sigma_{\k}(\e_n)
 = T\sum_{\e_{n'}\k'} \lim_{\p\rightarrow 0}
 \Gamma(\k\e_n;\k+\p,\e_n|\k'+\p,\e_{n'};\k'\e_{n'}) \cdot
 G_{\k'+\p}(\e_{n'}) G_{\k'}(\e_{n'}) \cdot 
 \e_{n'} {\vec v}_{\k'}^0 .
 \label{eqn:Ap-GWI3}
\end{eqnarray}

After the analytic continuation of eq.(\ref{eqn:Ap-GWI3}),
we find that
\begin{eqnarray}
% {\vec \nabla}_k \left(\e_\k^0 + \Sigma_{\k}^R(\e) \right) 
% \cdot \e
 {\vec \nabla}_k \Sigma_{\k}^{\rm R}(\e) \cdot \e = 
%{\vec v}_\k^0\e + 
\sum_{\k',i=1,3} \int \frac{d\e'}{4\pi i}
 {\cal T}^{1i}(\k\e,\k'\e') g_{\k'}^{(i)}(\e') 
 \cdot {\vec v}_{\k'}^0 \e' ,
 \label{eqn:Ap-GWI4}
\end{eqnarray}
where the four-point vertex ${\cal T}^{ij}$
is introduced by Eliashberg in ref. 
 \cite{Eliashberg}, 
and $g_{\k}^{(i)}(\e)$ is introduced in \S II.
Thus,
the Ward identity, eq.(\ref{eqn:Ap-GWI4}),
is derived in terms of the diagrammatic technique. 
Equation (\ref{eqn:Ap-GWI4})
is equivalent to eq.(\ref{eqn:qi2}),
which is expressed in terms of 
the zero temperature perturbation method.

We note that
``the usual Ward identity'' related to
the charge conservation law is given by
 \cite{AGD,Nozieres}
\begin{eqnarray}
 {\vec \nabla}_k \Sigma_{\k}^{\rm R}(\e) 
 = \sum_{\k',i=1,3} \int \frac{d\e'}{4\pi i}
 {\cal T}^{1i}(\k\e,\k'\e') g_{\k'}^{(i)}(\e') {\vec v}_{\k'}^0 .
 \label{eqn:Ap-GWI5}
\end{eqnarray}
According to eqs. (\ref{eqn:Ap-GWI4}) and (\ref{eqn:Ap-GWI5}),
\begin{eqnarray}
\sum_{\k',i=1,3} \int \frac{d\e'}{4\pi i}
 {\cal T}^{1i}(\k\e,\k'\e') g_{\k'}^{(i)}(\e') 
 \cdot {\vec v}_{\k'}^0 \e'
 \ = \ \e \cdot \sum_{\k,i=1,3} \int \frac{d\e'}{4\pi i}
 {\cal T}^{1i}(\k\e,\k'\e') g_{\k'}^{(i)}(\e') {\vec v}_{\k'}^0 .
\end{eqnarray}
%

%%%%%%%%%%%%%%%%%%%%
% references
%%%%%%%%%%%%%%%%%%%%

%\end{multicols}{2}
\end{document}